\documentclass[letterpaper,11pt]{article}
\usepackage{mathpazo}
\usepackage{microtype}
\usepackage[utf8]{inputenc}
\usepackage{graphicx,fullpage,paralist}
\usepackage{amsmath, amssymb, amsthm}
\usepackage{comment,hyperref}
\usepackage{thmtools}
\usepackage{tikz}
\usepackage{pgfplots}
\usepackage{varwidth}
\pgfplotsset{compat=1.10}
\usepgfplotslibrary{fillbetween}
\usetikzlibrary{backgrounds}
\usetikzlibrary{patterns}
\usepackage{enumitem}
\usepackage{geometry}
\usepackage{array}
\usepackage{multirow}
\usepackage{stmaryrd}
\usepackage[font=small]{caption}
\usepackage{bbm}
\newcommand{\calK}{\mathcal{K}}
\newcommand{\calB}{\mathcal{B}}

\newcommand{\calL}{\mathcal{L}}

\newcommand{\calA}{\mathcal{A}}

\newcommand{\RR}{\mathbb{R}}

\newcommand{\ZZ}{\mathbb{Z}}

\newcommand{\PMD}{\mathsf{PMD}}

\newtheorem{theorem}{Theorem}
\newtheorem{lemma}{Lemma}

\newtheorem{definition}{Definition}

\newtheorem{proposition}{Proposition}
\newtheorem{problem}{Problem}

\usepackage[textsize=tiny,textwidth=2cm]{todonotes}

\usepackage{multirow}
\usepackage[noend]{algpseudocode}
\usepackage{algorithm,algorithmicx}
\usepackage{xcolor}

\newlength{\algofontsize}
\hypersetup{
	colorlinks,
	linkcolor={red!50!black},
	citecolor={blue!50!black},
	urlcolor={blue!80!black}
}

\usepackage{tabularx}
\newcommand{\multiline}[1]{%
  \begin{tabularx}{\dimexpr\linewidth-\ALG@thistlm}[t]{@{}X@{}}
    #1
  \end{tabularx}
}

\begin{document}
	\algrenewcommand\algorithmicrequire{\textbf{Input:}}
	\algrenewcommand\algorithmicensure{\textbf{Output:}}
	
	\title{Preserving Diversity when Partitioning:\\ A Geometric Approach
%	 \thanks{This work was partially supported by blablabla}
	 }
	
	\author{Sebastian Perez-Salazar
	\thanks{School of Industrial and Systems Engineering, Georgia Institute of Technology, USA. {\tt sperez@gatech.edu}}
	\and 	Alfredo Torrico		
	\thanks{CERC Data Science, Polytechnique Montr\'eal, Canada. {\tt alfredo.torrico-palacios@polymtl.ca}}
	\and Victor Verdugo
	\thanks{Institute of Engineering Sciences, Universidad de O'Higgins, Chile. {\tt victor.verdugo@uoh.cl}}
	}
\date{\vspace{-1em}}

\maketitle
\begin{abstract}
Diversity plays a crucial role in multiple contexts such as team formation, representation of minority groups and generally when allocating resources fairly. 
Given a community composed by individuals of different types, we study the problem of partitioning this community such that the global diversity is preserved as much as possible in each subgroup. 
We consider the diversity metric introduced by Simpson in his influential work that, roughly speaking, corresponds to the inverse probability that two individuals are from the same type when taken uniformly at random, with replacement, from the community of interest.
We provide a novel perspective by reinterpreting this quantity in geometric terms. 
We characterize the instances in which the optimal partition exactly preserves the global diversity in each subgroup. 
When this is not possible, we provide an efficient polynomial-time algorithm that outputs an optimal partition for the problem with two types. 
Finally, we discuss further challenges and open questions for the problem that considers more than two types.
\end{abstract}
\thispagestyle{empty}
\newpage

\section{Introduction}\label{sec:introduction}

Diversity is a complex and multidimensional concept that is regularly used as a way to summarize the structure of a community. Addressing diversity concerns in decision-making tasks and socio-technical systems at large has become an imperative goal to achieve fairness and equity. Societal issues around emerging technologies and technical artifacts, such as datasets, have recently motivated the machine learning and artificial intelligence communities to study the notion of diversity and the related concepts of inclusion and representation~\cite{mitchell2020diversity,chasalow2021representativeness,chi2021reconfiguring,drosou2017diversity}. However, towards this goal, we face an immediate challenge: how to measure diversity. The chosen metric highly depends on the problem and the decision-maker's objectives. As Baumg\"artner \cite{baumgrtner2006measuring} observes: the choice of the diversity metric is conditioned to some extent on the aspects of diversity that the decision-maker considers more important. Numerous technical notions of diversity have been proposed in the literature such as  those based in abundance and similarity---common in biology and ecology---e.g. \cite{simpson1949measurement,macarthur1965patterns, baselga2007multiple,jost2009mismeasuring,leinster2012measuring}, geometry or distance-based, e.g. \cite{deshpande2010efficient,gong2014diverse,celis2018fair,samadi_etal18} and more recent ones that incorporate the concept of inclusion \cite{mitchell2020diversity}.

Diversity indices based on abundance aim to gauge the variety or heterogeneity of a community, without focusing on the specific attributes of each individual. Particularly, there is a long-standing consensus in ecology on recommending the usage of the Hill numbers \cite{hill1973diversity,heip1998indices,chao_etal14,daly2018ecological}, which satisfy key mathematical axioms and possess other desired properties \cite{daly2018ecological}. This class of indices includes the well-known \emph{Simpson dominance index}~\cite{simpson1949measurement}. Roughly speaking, this index represents the inverse of the probability that two individuals taken uniformly at random, with replacement, from the community of interest, are from the same type. Formally, consider a community with $r$ types of individuals where each type $i\in[r]:=\{1,\ldots,r\}$ has a relative abundance $p_i\in(0,1)$ with $\sum_{i\in[r]}p_i = 1$. Then, the Simpson dominance index corresponds to $1/\sum_{i\in[r]}p^2_i$. %Note that the index value always lies in the interval $[1,r]$. 

The Simpson dominance index, as other Hill numbers, weighs more on common types than rarer ones. The index reaches its maximum value $r$ when all types have equal relative abundance (evenness), i.e., $p_i = 1/r$ for all $i\in[r]$. 
%In other words, the effective number of different types equals the total number of types in the community. 
On the other hand, the index attains its minimum value of 1 when a single type has a relative abundance close to 1. Namely, there is some $i^{*}\in [r]$ with $p_i \approx 0$ for every $i\neq i^*$ and $p_{i^*}\approx 1$. The index's value always lies in the interval $[1,r]$ and uniquely depends on the abundance profile. 
Baumg\"artner \cite{baumgrtner2006measuring} provides the example in Table \ref{table:simpson_comparison} to compare the effects of the abundance profile in the value of the index. Note that subgroup $S_1$ has 4 types where each type is equally abundant. Similarly, subgroup $S_2$ has an index value of 5, however, it  is richer than $S_1$ since it has more types. Observe in subgroups $S_3$ and $S_4$ how the index decreases as the abundance of type 5 decreases. Subgroup $S_5$ shows that it is equally richer than $S_4$ but much less even. Finally, subgroup $S_6$ is richer than $S_5$ but almost equally even. %The Simpson dominance index captures the richness and evenness of the community. However, abundance-based indices do not take into account the attributes of each type. For example, if the community is composed by 3 types of individuals, they could be: (1) 3 different mammals or (2) 1 type of fruit, 1 type of vegetable and 1 type of meat. To address this, similarity-based adaptation have been considered CITE.
%More blabla on the properties of this index
\begin{table}[h!]
\centering
\begin{tabular}{|c|cccccc|}
\hline
Type & \multicolumn{6}{c|}{Relative abundance $p_i$ in subgroup $S$} \\
$i$ & $S_1$ & $S_2$ & $S_3$ & $S_4$ & $S_5$ & $S_6$ \\
\hline
$1$ &0.25&0.2&0.24&0.249& 0.50 & 0.50\\
$2$ &0.25&0.2&0.24&0.249& 0.30 & 0.30\\
$3$ &0.25&0.2&0.24&0.249& 0.10 & 0.10\\
$4$ &0.25&0.2&0.24&0.249& 0.07 & 0.07\\
$5$ & - &0.2&0.04&0.004& 0.03 & 0.01\\
$6$ & - & - & - & - & - & 0.01\\
$7$ & - & - & - & - & - & 0.01\\
\hline
Simpson index & 4.00 & 5.00 & 4.48 & 4.08& 3.42 & 3.53\\
\hline
\end{tabular}
\caption{Comparison for different Simpson dominance index values.}
\label{table:simpson_comparison}
\end{table}

In this work, we study the effects on diversity when partitioning a finite community of individuals into subgroups and we consider the Simpson dominance index as a diversity metric.
%, mainly because of its tractability.
Broadly speaking, the main question that we aim to address is the following:
\begin{quote}
\centering
\emph{
Given a positive integer value $k$, how do we divide a community into $k$ subgroups such that each subgroup's diversity is approximately as good as the global diversity?
}
\end{quote}
%This question can be interpreted as a fair division of indivisible objects CITE, where the value of each subgroup corresponds to its diversity.
Our main contributions in this work are the following: 
(1) to the best of our knowledge, we introduce the first model that incorporates diversity requirements in a partition problem;
(2) we give a geometric interpretation of each subgroup's diversity and its relationship with the global diversity;
(3) we characterize the instances in which the diversity of each subgroup is guaranteed to be at least as good as the global diversity;
(4) we provide a polynomial-time algorithm that outputs a partition of $k$ subgroups that preserve the global diversity as much as possible when the community is composed by 2 types;
(5) we discuss further challenges and open questions for the problem that considers more than 2 types.

%we investigate the problem of partitioning a set of individuals of a finite community into subgroups that exhibit a diversity comparable to the diversity of the global community. 
%
%maybe say something about fair division problem under the Simpson dominance metric.
%
%Motivated by these challenges, we investigate the problem of partitioning a set of individuals of a finite community into subgroups that exhibit a diversity comparable to the diversity of the original community. A finite community is inherently divided into types (e.g. genotypes) where each entity of the community belongs to one type. These types determine a global diversity of the community. Given an positive integer value $k$, we aim to divide the community into $k$ groups of entities in such a way that the diversity of these groups is comparable to the global diversity. 
%
%Our main framework is a resource constrained setting where blabla (explain in words the setting) and pose the main question (center and emphasized). Maybe say that it can be equivalently viewed as fair division of indivisible goods problem. Roughly state the contributions.

%In this work, we follow an approach inspired in ecological studies, and we measure diversity using the \emph{Simpson dominance index}~\cite{simpson1949measurement}. This index represents the inverse of the probability that two individuals taken uniformly at random from the group of interest (with replacement) are from the same type, that is, $1/\sum_{i=1}^rp_i^2$, where $p_i$ is the sampling probability of a type $i\in \{1,\ldots,r\}$. 
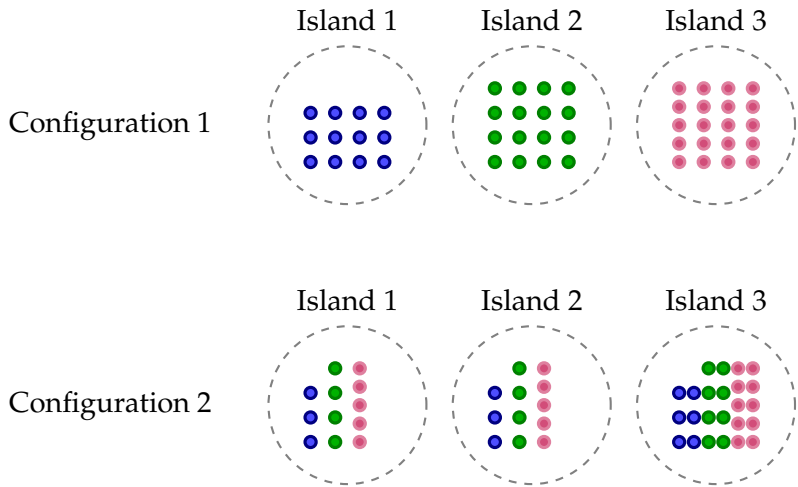
\begin{figure}[h!]
	\centering
	\begin{tikzpicture}[scale=0.7]
		
		\def\radius{1}
		\def\side{0.7}
		%islands
		\begin{scope}[xshift=-8cm]
			\node at (0,0) {Configuration 1 };
		\end{scope}
		
		\begin{scope}[xshift=-3.5cm]
			% island 1
			\node at (0,2) {Island 1};
			\draw[thick, dashed, gray] (0,0) circle ({1.5*\radius});
			\foreach \i in {0,...,3}
			\foreach \j in {0,...,2}
			{
				\pgfmathtruncatemacro{\x}{(2*\i - 3)}
				\pgfmathtruncatemacro{\y}{(2*\j - 3)}
				\draw[very thick, blue!50!black, fill=blue!70!white] (\x * \side/3, \y * \side/3) circle (3pt) node {};
			}
		\end{scope}

		% island 2
		\node at (0,2) {Island 2};
		\draw[thick, dashed, gray] (0,0) circle ({1.5*\radius});
		\foreach \i in {0,...,3}
		\foreach \j in {0,...,3}
		{
			\pgfmathtruncatemacro{\x}{(2*\i - 3)}
			\pgfmathtruncatemacro{\y}{(2*\j - 3)}
			\draw[very thick, green!50!black,fill=green!70!black] (\x * \side/3, \y * \side/3) circle (3pt) node {};
		}
		
		\begin{scope}[xshift=3.5cm]
			% island 1
			\node at (0,2) {Island 3};
			\draw[thick, dashed, gray] (0,0) circle ({1.5*\radius});
			\foreach \i in {0,...,3}
			\foreach \j in {0,...,4}
			{
				\pgfmathtruncatemacro{\x}{(2*\i - 3)}
				\pgfmathtruncatemacro{\y}{(\j - 2)}
				\draw[very thick, purple!50!white,fill=purple!70!white] (\x * \side/3, \y * \side/2) circle (3pt) node {};
			}
		\end{scope}
	\end{tikzpicture}
	
	~
%	\hrule
	~
	\\
	~
	
	\begin{tikzpicture}[scale=0.7]
		
		\def\radius{1}
		\def\side{0.7}
		%islands
		\begin{scope}[xshift=-8cm]
			\node at (0,0) {Configuration 2 };
		\end{scope}
		
		\begin{scope}[xshift=-3.5cm]
			% island 1
			\node at (0,2) {Island 1};
			\draw[thick, dashed, gray] (0,0) circle ({1.5*\radius});
			\foreach \j in {0,...,2}
			{
				\pgfmathtruncatemacro{\y}{(2*\j - 3)}
				\draw[very thick, blue!50!black, fill=blue!70!white] (-3 * \side/3, \y * \side/3) circle (3pt) node {};
			}
			\foreach \j in {0,...,3}
			{
				\pgfmathtruncatemacro{\y}{(2*\j - 3)}
				\draw[very thick, green!50!black,fill=green!70!black] (-1 * \side/3, \y * \side/3) circle (3pt) node {};
			}
			\foreach \j in {0,...,4}
			{
				\pgfmathtruncatemacro{\y}{(\j - 2)}
				\draw[very thick, purple!50!white,fill=purple!70!white] (1 * \side/3, \y * \side/2) circle (3pt) node {};
			}
		\end{scope}

		% island 2
		\begin{scope}[xshift=0cm]
			% island 1
			\node at (0,2) {Island 2};
			\draw[thick, dashed, gray] (0,0) circle ({1.5*\radius});
			\foreach \j in {0,...,2}
			{
				\pgfmathtruncatemacro{\y}{(2*\j - 3)}
				\draw[very thick, blue!50!black, fill=blue!70!white] (-3 * \side/3, \y * \side/3) circle (3pt) node {};
			}
			\foreach \j in {0,...,3}
			{
				\pgfmathtruncatemacro{\y}{(2*\j - 3)}
				\draw[very thick, green!50!black,fill=green!70!black] (-1 * \side/3, \y * \side/3) circle (3pt) node {};
			}
			\foreach \j in {0,...,4}
			{
				\pgfmathtruncatemacro{\y}{(\j - 2)}
				\draw[very thick, purple!50!white,fill=purple!70!white] (1 * \side/3, \y * \side/2) circle (3pt) node {};
			}
		\end{scope}
		
		\begin{scope}[xshift=3.5cm]
			% island 1
			\node at (0,2) {Island 3};
			\draw[thick, dashed, gray] (0,0) circle ({1.5*\radius});
			\foreach \i in {0,1}
			\foreach \j in {0,...,2}
			{
				\pgfmathtruncatemacro{\x}{(2*\i - 5)}
				\pgfmathtruncatemacro{\y}{(2*\j - 3)}
				\draw[very thick, blue!50!black, fill=blue!70!white] (\x * \side/5, \y * \side/3) circle (3pt) node {};
			}
		
			\foreach \i in {2,3}
			\foreach \j in {0,...,3}
			{
				\pgfmathtruncatemacro{\x}{(2*\i - 5)}
				\pgfmathtruncatemacro{\y}{(2*\j - 3)}
				\draw[very thick, green!50!black,fill=green!70!black] (\x * \side/5, \y * \side/3) circle (3pt) node {};
			}

			\foreach \i in {4,5}
			\foreach \j in {0,...,4}
			{
				\pgfmathtruncatemacro{\x}{(2*\i - 5)}
				\pgfmathtruncatemacro{\y}{(\j - 2)}
				\draw[very thick, purple!50!white,fill=purple!70!white] (\x * \side/5, \y * \side/2) circle (3pt) node {};
			}
		\end{scope}
	\end{tikzpicture}
	\caption{Two possible configurations of islands.}\label{fig:example_islands}
\end{figure}

\noindent {\bf Motivating example.} There are countless benefits and practical applications of finding diverse partitions. We offer this motivating example for concreteness~\cite{jost2009mismeasuring}. Here we use loosely the term diversity to refer to Simpson dominance index. Other metrics of diversity yield similar conclusions (see~\cite{daly2018ecological} for other metrics). We consider a population of $48$ entities with 3 types: $12$ blues, $16$ greens and $20$ pinks. The population is divided into three islands. In Figure~\ref{fig:example_islands}, we show two possible configurations of these divisions. The islands are represented by dashed circles. In configuration 1 (on the top), we have a homogeneous division of the population; in configuration 2 (on the bottom), we have a more heterogeneous division of the population. The global diversity of the population is $\gamma= 1/((12/48)^2 + (16/48)^2 + (20/48)^2) = 2.88 $. Suppose that some instantaneous natural catastrophe wipes out Island 1, leaving no survivors in that island. In configuration 1, after all blue types disappear, the new global diversity is $\gamma'= 1/((16/36)^2 + (20/36)^2 ) \approx 1.975$; hence, the diversity of the population decreases by $\approx 31.4 \%$. In configuration 2, the new diversity, after Island 1 disappears, is $\gamma'' = 1/((9/36)^2 + (12/36)^2 + (15/36)^2) = 2.88 = \gamma$; hence, there is no loss in diversity under the Simpson dominance index.
We can see from this example that ensuring diversity on each part (island) provides a more resilient configuration.

%\alfredo{explain problem with classical indices and usage of its associated effective numbers}

%\alfredo{solutions of SOCP are counter-intuitive, not necessarily balance (like the standard previous notions of diversity}

\subsection{Our Model}\label{sec:formulation}

We consider the problem of partitioning a community with $r\in\ZZ_+$ types into $k\in\ZZ_+$ groups or parts that preserve as much as possible the global diversity. Formally,
the input of the problem corresponds to a vector $b=(b_1,\ldots,b_r)\in \ZZ_+^r$, where the number $b_i$ denotes the amount of entities of type $i\in[r]$ in the community.
%; and a parameter $k\in \ZZ_+$ denoting the number of diverse groups to be formed. 
The output of the problem is a collection of vectors $x_1,\ldots,x_k\in \ZZ_+^r$ such that $\sum_{i=1}^k x_i = b$. We call such a collection a \emph{$k$-partition} of $b$.
% into $k$ parts.
The size of the community---the overall number of entities---corresponds to $n=\sum_{i=1}^r b_i$. Note that without loss of generality we can assume that $b_1\leq b_2\leq \ldots \leq b_r$. In this work, we consider the case when $k\leq b_1$, i.e., there exists a sufficient number of entities of each type for each part. 
%We assume that $k\leq n$ and, without loss of generality, we further assume that $b_1\leq b_2\leq \ldots \leq b_r$.
%Consider $r\in\ZZ_+$ different types and $b_i$ indivisible resources of each type $i\in[r]$. We denote by $\calX$ the ground set of $n := \sum_{i\in[r]}b_i$ resources and $b=(b_1,\ldots, b_r)$ the \emph{palette} or \emph{budget vector} of colors. We assume \alfredo{w.l.o.g. that $b_1\leq b_2\leq \cdots\leq b_r$ and } that balls of the same color are indistinguishable. Therefore, with a slight abuse of notation, we identify any subset $S\subseteq\calX$ with its allocation vector $(x_1,\ldots,x_r)$, where $x_i$ corresponds to the number of balls of color $i\in[r]$ in $S$. Finally, we denote the universe of feasible allocations by $\calA =\{x\in\ZZ^r_+: \ x_i\in\{0,\ldots,b_i\}, \ \forall i\in[r]\}$. 
Our goal is to form partitions $x_1,\ldots,x_k$ of $b$ that are as diverse as possible. 
%We use borrow notions from ecology and we use the \emph{Simpson diversity index}~\cite{simpson1949measurement} as a measure of diversity. 
%As mentioned earlier, the Simpson diversity index corresponds to the probability of simultaneously picking two entities of the same type in $S$ in a uniformly random experiment over the population (with replacement).
We measure diversity using the Simpson dominance index \cite{simpson1949measurement}. 
First, note that the relative abundance of type $i\in[r]$ is $p_i=b_i/\sum_{j=1}^rb_j = b_i/n$, so the global diversity corresponds to $n^2/\sum_{j=1}^rb_j^2$.
We now formalize the definition of this diversity measure for any vector $x\in\ZZ^r_+$.
%At the same time, we also define the \emph{Hill effective number} of $S$ as the multiplicative inverse of the Simpson diversity index. Formally, given a subset $S\subseteq \calX$,
%\alfredo{we assume that $k\leq b_1\leq \cdots\leq b_r$}
%\victor{Hay que modificar la def de $H$, estamos usando $H$ dado por L1 cuadrado dividido L2 cuadrado}
\begin{definition}
Given a vector $x\in \ZZ_+^r$, 
%the Simpson diversity index of $x$ is given by $H(x) =\left( \|x\|_2/\|x\|_1\right)^2$. 
the Simpson dominance index of $x$ is given by 
\[D(x) =\frac{\|x\|^2_1}{\|x\|^2_2},\] where $\|\cdot\|_1$ and $\|\cdot \|_2$ denote the $\ell_1$ norm and $\ell_2$ norm, respectively.
\end{definition}
In what follows, we refer to the Simpson dominance index of $x$, $D(x)$, just by \emph{diversity} of $x$, unless specified. 
The global diversity of a community determined by $b\in \ZZ_+^r$ is $D(b)$.
%To ease notation, we refer to $D(b)$, the diversity of the population, by $\gamma(b)$, or simply by $\gamma$ if $b$ is clear from context\footnote{The term gamma diversity is often used in ecology literature.}. 
%We take an approach borrowed from approximation algorithms analysis and we divide the task of finding diverse partitions into two closely related tasks: feasibility and maximum representativity.
%\begin{comment}
%\begin{problem}[Feasibility]\label{prob:feasibility}
%	Given $b=(b_1,\ldots,b_r)\in \ZZ_+^r$, $\epsilon\in[0,1)$ and $k\in\ZZ_+$, does there exist a partition $x_1,\ldots, x_k\in \ZZ_+^r$ of $b$ such that $D(x_i)\geq (1-\varepsilon)\cdot\gamma$ for all $i\in[k]$? 
%\end{problem}
%
%Note that any partition of $b$ into $k$ parts is feasible for $\epsilon = 1$. If we get a positive answer for the feasibility problem for $\varepsilon$, then we know that the feasibility problem will get a positive answer for any $\varepsilon'>\varepsilon$. Also note that this is not an optimization problem. 
%\end{comment}
The first natural question that arises is the following: Can we form a $k$-partition such that each part \emph{completely} preserves the diversity of the entire community. Formally, we define the \emph{perfect partition problem} as follows:

\begin{problem}[Perfect Partition]\label{prob:separation}
	Given $b=(b_1,\ldots,b_r)\in \ZZ_+^r$ and $k\in\ZZ_+$, does there exist a partition $x_1,\ldots, x_k\in \ZZ_+^r$ of $b$ that satisfies $D(x_i)\geq D(b)$ for all $i\in [k]$?
\end{problem}

When the previous problem has a negative answer, we address the following question: how close can we get to the global diversity? In other words,
%We now formally describe the problem we address in this work.
our goal is to compute a partition such that the diversity of each part is as close as possible to the global diversity. Formally, we define the \emph{problem of maximin diversity} as follows,

\begin{problem}[Partition of Maximin Diversity ($\PMD$)]\label{prob:optimal}
	Given $b=(b_1,\ldots,b_r)\in \ZZ_+^r$ and $k\in\ZZ_+$, what is the minimum $\varepsilon\geq0$ such that there exists a partition $x_1,\ldots, x_k\in \ZZ_+^r$ of $b$ that satisfies $D(x_i)\geq (1-\varepsilon)\cdot D(b)$ for every $i\in [k]$? We denote this value by $\varepsilon(b,k)$.
\end{problem}

Note that Problem \ref{prob:optimal} can be equivalently stated as
\begin{equation}\label{eq:maximin}
\max\Big\{\min\{D(x_1),\ldots,D(x_k)\}:x\in \ZZ_+^k\text{ and }x_1+\cdots+x_k=b\Big\},
\end{equation}
which can be seen as an analogous of the fair division problem \cite{budish2011combinatorial,lipton2004approximately,bouveret2016fair}.\\
%\alfredo{add the min representation of the problem as a comment
%Problem~\ref{prob:feasibility} and Problem~\ref{prob:optimal} are naturally connected. Answering Problem~\ref{prob:optimal} gives us an answer for Problem~\ref{prob:feasibility}; while answering Problem~\ref{prob:feasibility} gives us a method to answer Problem~\ref{prob:optimal} via bisection over $\varepsilon>0$. 
%On top of these two problems, we also consider a weaker question---answerable by any of the previous problems---but with its own merit. 
%The separation problem corresponds to the feasibility problem with parameter $\varepsilon=0$. The \emph{separation} term in the problem's name comes from the fact that we are separating the $\varepsilon$ (the loss of diversity) from $0$. The separation problem address the question of whether there is a way to partition the population into groups with the same (or better) diversity as the original population.
%\sebastian{To add a bunch of examples showing challenges}

\noindent {\bf Tentative approaches.} To give some insights behind the difficulty of finding these optimal partitions, consider the natural approach of distributing items in a proportional manner by the so-called \emph{balanced} solutions or \emph{round-robin} procedures. 
The idea is to balance types over different parts in an iterative manner. 
We show an example for which some balanced solutions are not always optimal. 
Consider $b=(6,14,21)$ and $k=2$. We have $D(b)=1681/673\approx2.497$. A balanced solution is $x_1=(3,7,10)$ and $x_2=(3,7,11)$. This solution ensures that the diversity of each part is $\varepsilon \approx 0.013$ away from the global diversity. However, the solution: $x_1'=(3,6,10)$ and $x_2'=(3,8,11)$ ensures that the diversity of each part is $\varepsilon\approx 0.003$ away from the global, almost four times smaller than the balanced solution.

%These are solutions typically formed by round-robin procedures. 
In other example, consider $b=(6,15,21)$ and $k=2$. We have $D(b)= 98/39\approx 2.512$. 
One possible partition of $b$ into $2$ parts is: $x_1= (3,7,10)$ and $x_2=(3,8,11)$ obtained by setting the entries of $x_1$ to be the floor of $b_i/2$ with $i\in \{1,2,3\}$; the rest is given to $x_2$. It is easy to check that $D(x_2)/D(b) \approx 0.992$. On the other hand, the partition: $x_1'=(2,5,7)$ and $x_2'=(4,10,14)$ holds $D(x_1')=D(x_2')= D(b)$, that is, we can match exactly the global diversity with this optimal partition $(x_1',x_2')$.\\
%Therefore we have feasibility with $\varepsilon=0$ here and the separation problems holds.

\noindent {\bf Notation.} In the remainder of the manuscript, we follow the following notation. We denote by $\mathbf{1}$ the all-1 vector in $\RR^r$, $\langle \cdot, \cdot \rangle$ the standard euclidean inner product, $\theta_x$ the angle formed between vectors $x\in\RR^r$ and $\mathbf{1}$. Also, denote by $\gcd(b)= \gcd(b_1,\ldots,b_r)$ the greatest common divisor of the elements $b_1,\ldots,b_r$. 

\subsection{Our Results}\label{sec:contributions}

Our first result fully answer Problem~\ref{prob:separation}. We show that a $k$-partition with each part as diverse as the the total population can be achieved if and only if $k$ is at most the greatest common divisor of $b=(b_1,\ldots,b_r)$.
%\begin{theorem}\label{thm:main_1}
%	A perfect partition $x_1,\ldots,x_k\in \ZZ_+^r$ of $b$ exists if and only if $k\leq \gcd(b_1,\ldots,b_r)$.
%\end{theorem} 
An immediate implications of this result is that diversity deteriorates when $k$ is larger than the greatest common divisor. 
The analysis and proof of this result can be found in Section \ref{sec:perfect-partition}.
%The proof of the theorem relies

Our next result answers the $\PMD$ problem (Problem~\ref{prob:optimal}) for the case of two types, $r=2$. We present an algorithm that outputs a partition $(x_1,\ldots,x_k)$ achieving $\varepsilon(b,k)$. Moreover, the implementation of this algorithm is $\mathcal{O}(k\log^2 \max\{ b_1,b_2 \})$, which is polynomial in the input $(b_1,b_2)$ and polynomial in the output of size $k$. 
%\begin{theorem}\label{thm:main_2}
%	There is a polynomial time algorithm that solves exactly Problem~\ref{prob:optimal} for $r=2$.
%\end{theorem}
More specifically, we show that 
Problem~\ref{prob:optimal} has the following geometric interpretation: Find a piecewise linear curve that joins $0$ and $b$ with $k-1$ breakpoints and such that each part has a diversity as good as $b$ (see Figure~\ref{fig:example_two_colors}). A simple calculation shows that the diversity of a vector $x$ is proportional to $(\cos \theta_x)^2$, where $\theta_x$ is the angle formed between $x$ and the vector $\mathbf{1}=(1,\ldots,1)$. Thus, we aim to find a piecewise approximation of the line $\mathcal{L}=\{ t \cdot b: t\in [0,1]  \}$ where each segment forms angles as close as possible with the vector $\mathbf{1}$. 

As a warm up, we first present the analysis for $k=2$ in Section \ref{sec:warm_up_optimal}. 
%We characterize the solution of $k=2$ as the problem of finding the closest integer vector $x\neq 0, b$ above the line $\mathcal{L}$. The two optimal parts are then given by $x$ and $b-x$. If $\gcd(b_1,b_2) > 1$, this claim is trivial as it can be answered using Theorem~\ref{thm:main_1}. For the case $\gcd(b_1,b_2)=1$ we characterize the distances of the closest integral points $\neq 0,b$ to line $\calL$. We show that they are unique multiples of a constant, and that the closest one minimizes their slopes, hence their angle with the vector $\mathbf{1}$. Relating distance with angle is a nontrivial fact
For the case $k\geq 3$, we show a structural result that characterizes the optimal solutions with $k$ parts under mild assumptions.
% (Lemma~\ref{lem:optimal_k_parts}). 
Then, by decreasing appropriately $b_1$ and $b_2$, we can find an instance where we can solve the $k$ partition problem and we implement this procedure recursively. 
%Our algorithms are inspired by the Extended Euclidean Algorithm.
The analysis can be found in Section \ref{sec:warm_up_optimal}. Finally, in Section \ref{sec:challenges}, we discuss the geometric challenges that the $\PMD$ problem poses in higher dimension and we propose two tentative approaches to find $2$-partitions. We also discuss the main open questions related to the complexity of the problem and the usage of other diversity metrics.

\subsection{Related Work}\label{sec:related}

One of the objectives of diversity in ecology is to gauge rare species in a population. There is a spectrum of viewpoints; on one side, rare species are the main focus; while on the other side, communities are important and only measuring common species matters. In his Nature's influential work~\cite{simpson1949measurement}, Simpson introduced a sample-driven metric of diversity based on abundance and richness of a population. The Simpson dominance index weighs heavily on rare species. Also, it has been generalized to the Hill numbers~\cite{hill1973diversity}, and more generally, it has been derived as special cases of entropy indices~\cite{keylock2005simpson}. These more general numbers allow practitioners to weighs on more common species. More modern metrics include similarities between different types or species~\cite{leinster2012measuring}. For additional indices and metrics of diversity, we refer the interested reader to~\cite{daly2018ecological}. In this work, we digress from the Simpson's index probabilistic viewpoint and we interpret this index as a geometric object on a high dimensional space. Closely related to our geometric approach is the cosine similarity~\cite{xia2015learning}.

%Diversity is an elusive concept though, so it is the inclusionof suchqualitativemeasure in machines that only learn from data.

Diversity has an essential role in many areas outside ecology. For instance, fairness in data summarization~\cite{celis2018fair,celis2020fair,huang2019stable}, where the goal is to select a small group of representative data that exhibits diversity in the feature space and is fair among sensitive features. A closely related approach is fair clustering~\cite{anegg2020technique,ghadiri2020fair,chierichetti2018fair,jia2020fair}. The majority of this literature has focused on producing clusters where no protected class is underrepresented. Our techniques could help provide new geometric insights in the design of fair and diverse outputs. There has also been a growing interest in building algorithms that are diverse in the sense of membership-aware~\cite{ahmed2017diverse,bandyapadhyay2019constant,agrawal18b}.

Another area where diversity has been extensively studied is recommendation systems~\cite{adomavicius2011improving,bradley2001improving,smyth2001similarity}. In this context, the main goal is to create better content-based recommendations by diversifying and not just rely on similar contents~\cite{mcsherry2002diversity,ziegler2005improving}. %We believe that finding partitions which preserve most of the diversity could be used to generate diverse recommendations.
As we mentioned in Section \ref{sec:formulation}, Problem \ref{prob:optimal} can be equivalently formulated as the fair division problem \eqref{eq:maximin}. This problem can be interpreted as the maximin guarantee~\cite{budish2011combinatorial} used in fair allocation of indivisible goods, which has been extensively studied~\cite{lipton2004approximately,bouveret2016fair,procaccia_wang14,kurokawa2016can,barman2017approximation,ghodsi_etal18,asadpour2010approximation}. 
The problem of splitting attributes in the construction of decision trees (see e.g. \cite{laber2018binary} and the references therein) is closely related to the problem we introduce in this work. The goal is to design splitting procedures that minimize the \emph{impurity} of the partitions, where impurity is measured with the Entropy or Gini metric. For example, given $x\in\ZZ^r_+$, the Gini impurity measure corresponds to $1-1/D(x)$. In particular, Laber et al. \cite{laber2018binary} design splitting procedures with constant approximation guarantees.
Finally, other notions of diversity and inclusion for subset selection tasks has been recently proposed in \cite{mitchell2020diversity}
%\alfredo{new work on diversity and inclusion \cite{mitchell2020diversity}}
%In our work, we aim to find a partition of $b$ in such a way that the least diverse part is as diverse as possible. 
%These guarantees have been extensively studied~\cite{}. See also~\cite{}. 

%\alfredo{donde se conecta esta bibliografia --- The problem of maximizing diversity has been studied in~\cite{leinster2009maximum,leinster2016maximizing}.}

%\section{Preliminaries}
\section{The Perfect Partition Problem}
\label{sec:perfect-partition}
%Let $b=(b_1,\ldots,b_r)$ be the vector of types and denote by $\gcd(b)= \gcd(b_1,\ldots,b_r)$ the \emph{greatest common divisor} of the elements $b_1,\ldots,b_r$. For $\ell, p$ nonzero integers, we denote by $\ell\mid p$ the property $\ell$ divides $p$. 

%In the next sections, we will we use the following classical result for the characterization of optimal solutions and the design of our algorithms.
%
%\begin{theorem}[Extended Euclidean Algorithm \alfredo{any cite? folklore?}]
%	Given $b_1,\ldots,b_r$ nonnegative integers and  $d=\gcd(b)$, there is an algorithm that obtains integers $k_1,\ldots,k_r$ such that
%	\[
%	k_1 b_1 + \cdots + k_r b_r = d
%	\]
%	in $\mathcal{O}(\log^2 (b_1+ \cdots b_r))$ time. 
%\end{theorem}

In this section, we characterize those instances in which there exists a partition that does not deteriorate the diversity of the whole population. First, we show the following properties of the Simpson diversity index.
%\alfredo{properties related to mathematical axioms cite}
%The following proposition shows properties of the function $H$ that are going to be useful in the analysis of partitioning problem. %Proof of these properties have been deferred to the Appendix
%\victor{modificar aca tambien la def de $H$}
\begin{proposition}
	\label{prop:technical-schur}
	The Simpson diversity index $D$ satisfies the following:
	\begin{enumerate}[label=(\alph*)]
		\item $D(\alpha x) = D(x)$ for any $\alpha\neq 0$. \label{schur-scaling}
%		\item $H$ is Schur-concave. \label{schur-schur} \sebastian{Usamos esto en algun lado?}
		\item For $x\geq 0$, $D(x)= r (\cos \theta_x)^2$ where $\theta_x$ is the angle formed between $x$ and $\mathbf{1}=(1,\ldots,1)\in \ZZ^r$. \label{schur-angle}
	\end{enumerate}
\end{proposition}

\begin{proof}
Recall that the Simpson diversity is defined as $D(x) =(\|x\|_1/\|x\|_2)^2$, from where we get directly that $D$ is invariant under scaling, which corresponds to Property \ref{schur-scaling}. For Property \ref{schur-angle}, observe that since $x$ is non-negative we have
\(\|x \|_1 = \langle x, \mathbf{1} \rangle = \|x\|_2\sqrt{r} \cos \theta_x.\)
\end{proof}

For other mathematical properties of this index, we refer the interested reader to \cite{daly2018ecological}.
Let us recall Problem \ref{prob:separation}: Given a vector of types $b$, does there exists a $k$-partition $x_1,\ldots, x_k$ such that $D(x_i) \geq D(b)$ for all $i\in[k]$? Observe, that this question is equivalent to characterize $\varepsilon(b,k)=0$ versus $\varepsilon(b,k)>0$ in Problem \ref{prob:optimal}. In the main result of this section, we show that is possible to solve Problem \ref{prob:separation} if, and only if, the number of parts do not exceed the greatest common divisor of $b$.

%\subsection{Characterizing the Instances where $\varepsilon(b,k)=0$}

%In other words, we present a characterization of the maximum representativity problem when $\varepsilon^*=0$. 
\begin{theorem}\label{lem:characterization_gcd_eps}
	For every $b=(b_1,\ldots,b_r)\in \ZZ_+^r$ and $k\in\ZZ_+$, there exists a $k$-partition $x_1,\ldots, x_k\in \ZZ_+^r$ of $b$ that satisfies $D(x_i)\geq D(b)$ for all $i\in [k]$ if, and only if, $k\leq \gcd(b)$.	
	%	MRP in $k$ parts has solution $\varepsilon_k^*=0$ if and only if $k\leq \gcd(b)$.
\end{theorem}

%We start by presenting a general result that gives a characterization when a budget vector $b=(b_1,\ldots,b_r)$ can be partitioned into $k$ parts, each part being at least as diverse as the budget vector. The following result characterizes the $\varepsilon(b,k)=0$ versus $\varepsilon(b,k)>0$. It states that we can find diverse partitions if and only if the number of partitions required do not exceed the gcd of $b$.
%%In other words, we present a characterization of the maximum representativity problem when $\varepsilon^*=0$. 
%\begin{lemma}\label{lem:characterization_gcd_eps}
%	For every $b=(b_1,\ldots,b_r)\in \ZZ_+^r$ and $k\in\ZZ_+$, there exists a partition $x_1,\ldots, x_k\in \ZZ_+^r$ of $b$ that satisfies $D(x_i)\geq D(b)$ for all $i\in [k]$ if and only if $k\leq \gcd(b)$.	
%	%	MRP in $k$ parts has solution $\varepsilon_k^*=0$ if and only if $k\leq \gcd(b)$.
%\end{lemma}

\begin{proof}	
	We prove both implications separately. Throughout the proof we denote $d=\gcd(b)$.	
	Let $k$ be the maximum integer such that there exists a partition $x_1,\ldots, x_k\in \ZZ_+^r$ of $b$ that satisfies $D(x_i)\geq D(b)$ for all $i\in [k]$.
	We aim to show that $k=d=\gcd(b)$. 
	%The result will follow immediately from this.
%	We first observe that since $d$ divides $b_i$ for every $i\in [d]$, then $b/d$ is an integral vector. Moreover, $D(b/d) = D(b)$ by Proposition~\ref{prop:technical-schur}. Thus, $b/d$ is a feasible solution of the maximum representativity problem with $\varepsilon=0$ and $\varepsilon(b,d)=0$. Therefore, $k\geq d$. 
First, observe that since $d$ divides $b_i$ for every $i\in [r]$, then $b/d$ is an integral vector. Moreover, $D(b/d) = D(b)$ by Property \ref{schur-scaling} in Proposition~\ref{prop:technical-schur}. Thus, $b/d$ is a feasible solution of the $\PMD$ problem with $\varepsilon=0$ and $\varepsilon(b,d)=0$. Therefore, $k\geq d$. 
	Now, suppose by contradiction that $k\geq d+1$. Take any solution $x_1,\ldots,x_k$ of the PMD problem with $k$ parts and $\varepsilon=0$, that exists by the choice of $k$. Then we have $D(x_i) \geq D(b)$ for all $i\in [k]$. 
	We claim that the vectors $x_1,\ldots,x_k$ are aligned, that is, for every $j\in \{2,3,\ldots,k\}$ there exists a positive $\alpha_{j}$ such that $x_j= \alpha_{j} x_1$. 
	Indeed, if the vectors are not aligned, we have that 
	\begin{align*}
		\| b\|_1  = \sum_{i=1}^k \|x_{i}\|_1 \geq \sum_{i=1}^k \sqrt{D(b)} \|x_{i}\|_2  > \sqrt{D(b)}\cdot \Big\| \sum_{i=1}^k x_{i}\Big\|_2= \sqrt{D(b)}\|b\|_2,
	\end{align*}
	where the first inequality is a consequence of $D(x_i) \geq D(b)$ for all $i\in [k]$ and the strict inequality holds by the triangle inequality, which is strict since the vectors $x_1,\ldots,x_k$ are not aligned.
	The above chain of inequalities implies that  $D(b)<\|b\|_1^2/\|b\|_2^2$, which is a contradiction since this is an equality. 
	We conclude that the vectors $x_1,\ldots,x_k$ are aligned, and consequently, we have for each $j\in \{2,\ldots,k\}$ that $x_{ j} = \tilde{\alpha}_j b$ where $\tilde \alpha_j = \alpha_j/(\sum_{j=2}^{k} \alpha_j+1)$ and let $\beta_j,\delta_j$ coprime such that $\tilde{\alpha}_j = \beta_j/\delta_j$. From this, we note that for any $i\in[r]$,
	$\beta_j b_{i} = \delta_j x_{ij}$
	and so $\delta_j$ divides $b_{i}$ for each $j\in \{1,\ldots,k\}$, since $\beta_j$ and $\delta_j$ are coprime. This shows that, for each $j\in \{1,\ldots,k\}$, $\delta_j$ is a common divisor of $b_1,\ldots,b_r$ and therefore $\delta_j\leq d$. 
	We deduce from here that $\tilde\alpha_j = \beta_j/\delta_j \geq 1/d$. Then, 
	\[b = \sum_{j=1}^k x_j = \sum_{j=1}^k \tilde \alpha_j b \geq \frac{k}{d} b,\] 
	which implies $k\leq d$.
	Let us prove the opposite implication. Consider an integer value $k\le d$ and let $x_j=b/d$ for all $j\in \{1,\ldots,d\}$. As in the previous part, we have $D(x_i)= D(b)$. Now define $x_{j}' = x_{j} = b/d$ for every $j\in \{1,\ldots, k-1\}$ and 
	\[x_{k}'= b-\sum_{j=1}^{k-1} x_{j} = (d-k+1) \frac{b}{d}.\] 
	Then, by Proposition \ref{prop:technical-schur} \ref{schur-scaling} we have that $D(x_{j}')= D(b)$ for every $j\leq k$. This finishes the proof.
\end{proof}

\section{The Partition of Maximin Diversity for Two Types}\label{sec:warm_up_optimal}

\begin{figure}[h!]
	\centering
	\begin{tikzpicture}[scale=0.75]

		\foreach \x in {0,...,5}
		\foreach \y in {0,...,7} 
		\node [circle,fill=black,scale=0.23]  (\x\y) at (1*\x,1*\y) {}; 
		
		\node (-x) at (-0.5,0) {};
		\node (x) at (5.5,0) {};
		\node (y) at (0,7.5) {};
		\node (-y) at (0,-0.5) {};
		
		\node(b1) at (5,0) [scale=0.1] {};
		\node(b2) at (0,7) [scale=0.1] {};
		\node(b) at (5,7) [scale=0.3, fill=black, circle] {};
		\node(origin) at (0,0) [scale=0.1] {};

		\node(z0) at (1,2) [scale=0.4, fill=blue, circle] {};
		\node(z) at (3,5) [scale=0.4, fill=blue, circle] {};

		\path[draw=blue,every node/.style={sloped,anchor=south,auto=false}]
		(origin) edge node {} (z0)
		(z0) edge node {} (z)
		(z) edge node {} (b)
		;
		
		\path[draw=black,every node/.style={sloped,anchor=north,auto=false}]
		(b) edge node {} (origin)
		;
		
		%labels
		\node at (b) [right] {$b$};
		\node at (z0) [left] {$x_1$};
		\node at (z) [above left] {$x_1+x_2$};
		\node (origen) at (-0.3,-0.3) {$0$};
		
		\path[dashed,thin,draw=gray]
		(b1) edge node {} (b)
		(b) edge node {} (b2)
		;
		
		%		\path[red,thin]
		%		(z) edge node {} (3,4.2)
		%		(z) edge node {} (3.57,5)
		%		(3.57,5) edge node {} (3,4.2)
		%		;
		
		%axis
		\path[draw,thick,<->] 
		(-x) edge node {} (x)
		(y) edge node {} (-y);

	\end{tikzpicture}
	\caption{Lattice representation of a partition of $b=(5,7)$ into three parts. 
		The partition is given by $x_1=(1,2)$, $x_{2} = (2,3)$ and $x_{3}= (2,2)$. 
		The breakpoints correspond to $x_1=(1,2)$ and $x_1+x_2=(3,5)$.}
	\label{fig:example_two_colors}
\end{figure}
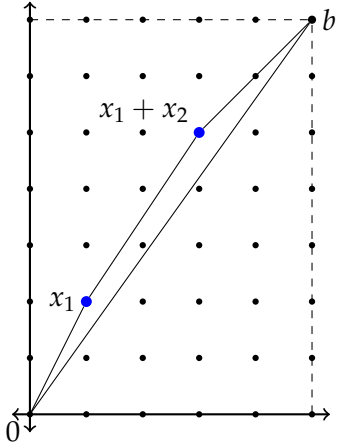

In this section, we present optimal algorithms for the $\PMD$ problem when the number of types equals two.
To illustrate the challenges of our problem, consider a geometric interpretation of the resources in the two dimensional plane. Represent each part $x_1,\ldots,x_k$ of the budget $b$ as points in the two dimensional integer lattice. Let $\calL=\{t b: \ t \in [0,1]\}$ be the segment that joins the origin and $b$. A $k$-partition can be visualized as a linear piecewise approximation of the line segment $\calL$ with $k-1$ breakpoints given by $\sum_{i=i}^{j}x_i$ for $j\in \{1,\ldots,k-1\}$ (Figure~\ref{fig:example_two_colors}). Since we are assuming $b_1\leq b_2$, intuitively, points lyings above the line $\calL$ will define the diversity of the partition. This is because the vector $\mathbf{1}$ lies below the line $\calL$ and $D(x)=2(\cos \theta_x)^2$ (Proposition~\ref{prop:technical-schur}), so if $\theta_x$ increases, then $D(x)$ decreases.

We show that the optimal solution is characterized by the $k$ closest points to the line $\calL$ that lie above the line. It is worth remarking that this is a non-trivial fact. For instance, a close point to the line that is also close to the origin could exhibit a larger angle to $\mathbf{1}$ than a point that is further from the origin and that is also further from $\calL$. In fact, we show that the distance of points of the form $(i, \lceil b_2 i /b_1 \rceil )$ with $i\in\{1,\ldots,b_1-1\}$ to the line are unique integral multiples of $1/b_1$. This allows us to show that the closest points to the line are also minimizing in terms of the angles formed with the vector $\mathbf{1}$. A naive implementation computes all the $b_1-1$ values and selects the best $k$ ones. We show how to improve this polylogarithmically in $b_1+b_2$ by using some basics of remainders. 

First, and for the ease of explanation, we present formally our ideas for the case $k=2$. We present an algorithm based on computing the closest point above the line $\calL$. In the second part, we analyze the general case of $k$-partition for two types. We build upon the results for $2$-partition and we show its correctness.

%\subsection{The Case of Two Types and Two Parts ($r=k=2$)}
\subsection{Warm-up: Analysis for $2$-Partitions}

In this section, we present a deterministic algorithm with complexity $\mathcal{O}( \log^2 (\max\{ b_1, b_2 \}) )$. 
%For the rest of the section, we are going to assume that $\gcd(b_1,b_2)=1$ since otherwise we know that we can easily solve the problem via Lemma~\ref{lem:characterization_gcd_eps}. We also assume that $2\leq b_1 \leq b_2$. Since the case $b_1=1$ is easy. 
%\sebastian{Explain why $k\leq b_i$ for all $i$}
Let $\mathcal{A}=\{ x \in \ZZ_+^2 : 0\leq x \leq b ,  1 \leq \| x\|_1 \leq \|b\|_1-1 \}$ be the set of all non-trivial feasible allocations for the first part in the division. 
Note that if $x\in\calA$ is the allocation for the first part, then the allocation for the second part corresponds to $b-x$, which also belongs to $\calA$. Hence, any $2$-partition of the $\PMD$ problem can be represented as $(x,b-x)$ where $x\in \calA$.

%Naturally, if $x$ is the allocation for part 1, the allocation for part $2$ equals to $b-x$; hence $x, b-x$ is a partition of $b$ in $2$ parts for every $x\in \calA$. 
%Note that if $x\in \mathcal{A}$, then $b-x\in \mathcal{A}$.

Observe that the lattice representation of any partition for $b$ is given by a linear piecewise approximation of the line segment $\calL$ with one breakpoint $x\in \mathcal{A}$. The key idea behind our algorithm is to be as close as possible to $b$, while being integral at the same time. This idea is inspired by our result for the separation problem in Theorem~\ref{lem:characterization_gcd_eps}. Our approach is formally presented in Algorithm~\ref{alg:two_colors_two_parts_algorithms}. Intuitively, our algorithm is searching over the values $i\in \{1,\ldots,b_1-1\}$ and selecting the closest integer point above the point $b_2 i/b_1$ in the line segment $\calL$. 
%This performed naively results in an algorithm with time $\mathcal{O}(\min\{b_1,b_2\})$. 
Algorithm~\ref{alg:two_colors_two_parts_algorithms} can be implemented using the well-known Extended Euclidean algorithm (Theorem~\ref{thm:extended_euclid}) in time $\mathcal{O}(\log^2 (\max\{ b_1, b_2 \}))$; a proof of this result can be found in \cite{cormen2009introduction}.

\begin{theorem}[\cite{cormen2009introduction}]\label{thm:extended_euclid}
	Given $b_1,\ldots,b_r$ non-negative integers and  $d=\gcd(b)$, there is an algorithm that obtains integers $k_1,\ldots,k_r$ such that
	\(
	k_1 b_1 + \cdots + k_r b_r = d
	\)
	in $\mathcal{O}(\log^2 (b_1+ \cdots + b_r))$ time. 
\end{theorem}

\begin{algorithm}[h!]
%\SetAlgoNoLine
	\caption{Two Types and Two Parts}\label{alg:two_colors_two_parts_algorithms}
	\begin{algorithmic}[1]
		\Require{Budget $b = (b_1,b_2)$ with $b_1\leq b_2$.}
		\Ensure{A partition $(x,b-x)$ of maximin diversity.}
		\State If $d=\gcd(b_1,b_2)\geq 2$ return the partition $(x,b-x)=(b/d,b-b/d)$\;
		\State If $\gcd(b_1,b_2)=1$, compute $\tau= b_2 \pmod {b_1}$ and
		let $i^*\in \{1,\ldots,b_1-1\}$ be such that $\tau \cdot i^* = b_1 - 1 \pmod{ b_1}$. 
		%for $i=1,\ldots,b_1-1$.
		\State Return the partition $(x,b-x)$ given by $x=(i^*, \lceil b_2 i^*/b_1 \rceil )$.
%		\EndIf
%		\Return $(x,y)$. 
	\end{algorithmic}
\end{algorithm}

%\alfredo{somewhere? -- Recall that we can assume w.l.o.g. that $b_1\leq b_2$.}

%Therefore, the time complexity follows. 
In the following, we show the main result of this section: the correctness of Algorithm~\ref{alg:two_colors_two_parts_algorithms}. Note that we only need to analyze the case when $\gcd(b_1,b_2) = 1$, since the other one was shown in Theorem~\ref{lem:characterization_gcd_eps}. 

\begin{theorem}
\label{thm:two-two}	
	For every budget $b=(b_1,b_2)$ such that $\gcd(b_1,b_2)=1$ and $b_1\le b_2$, the 2-partition $(x,b-x)$ computed by Algorithm~\ref{alg:two_colors_two_parts_algorithms} solves the $\PMD$ problem with two parts. 
	The algorithm runs in time $\mathcal{O}(\log^2 \max\{ b_1,b_2 \})$.
%	In particular, in this case we have that $\varepsilon(b,2) = 1 - \frac{D(x)}{D(b)}$.
\end{theorem}

To prove this theorem, we follow the next steps.
Note that the line segment $\calL$ divides the region $\mathcal{A}$ into two symmetric parts, $\mathcal{A}_+=\{y\in \calA:y_2\ge (b_2/b_1) y_1\}$ and $\mathcal{A}_-=\{y\in \calA:y_2\le (b_2/b_1)y_1\}$; above and below the line segment, respectively. 
In particular, we can assume that the upper part of the region $\mathcal{A}$ contains a solution of the $\PMD$ problem. Therefore, solving the $\PMD$ problem for two parts is equivalent in this case to the problem $\min_{y\in \calA_+}(1-D(y)/D(b))$. We then characterize the value $\varepsilon(b,2)$ as an equivalent optimization problem over $\calA_+$. Finally, we prove that the solution computed by Algorithm~\ref{alg:two_colors_two_parts_algorithms} solves this optimization problem; hence it solves the $\PMD$ problem. %In what follows we formalize these steps.

\begin{proposition}
	Let $(x,b-x)$ be an optimal solution of the $\PMD$ problem with two parts, such that $x\in \calA_+$.
	Then, we have $\varepsilon(b,2) = \min_{y\in \mathcal{A}_+}(1 - D(y)/D(b))=1 - D(x)/D(b)$. 
\end{proposition}

\begin{proof}
	Since $(x,b-x)$ is an optimal solution for the $\PMD$ problem, we have	that $\min\{D(x),D(b-x)\}= (1 - \varepsilon(b,2)) D(b)$ and
	therefore $\varepsilon(b,2) = 1 - \min\{D(x),D(b-x)\}/D(b)$. 
	On the other hand, by Proposition \ref{prop:technical-schur} \ref{schur-angle} we have that 
	$D(x) = 2 (\cos \theta_x)^2$
	where $\theta_x$ is the angle formed between $x$ and the vector $\mathbf{1}$. Similarly, we get $D(b-x) = 2 (\cos \theta_{b-x})^2$ and $D(b) = 2 (\cos \theta_b)^2$. 
	Then we have \[\varepsilon(b,2) = 1 - \min\Big\{\left(\cos \theta_x/\cos \theta_b \right)^2,\left(\cos \theta_{b-x}/\cos \theta_b\right)^2\Big\}.\]
	Since $x\in \calA_+$ while $b-x\in \calA_-$, { we have that $\theta_x > \theta_{b-x}$}. 
	Cosine is a decreasing function in $[0,\pi/2]$ and therefore we conclude that $\varepsilon(b,2) = 1 - \left(\cos \theta_{x}/\cos \theta_b\right)^2 = 1-D(x)/D(b)$.
\end{proof}

We refer by the \emph{vertical distance from the point $(i,j)\in \ZZ_+^2$ to the line segment $\calL$} to the quantity $|j - b_2 i/b_1|$. By abusing notation, we will refer to this quantity by simply \emph{vertical distance}. The following proposition states that all vertical distances of points $(i, \lceil b_2 i/b_1\rceil)$, $i=1,\ldots,b_1-1$, to the segment $\calL$ are in one-to-one correspondence to the set $\left\{1/b_1, 2/b_1, \ldots, (b_1 - 1)/b_1 \right\} $. We will use this fact to argue that no two different point $(i,\lceil b_1 i / b_1\rceil)$ and $(j,\lceil b_1 j / b_1\rceil)$ have the same vertical distance. %Figure~\ref{fig:prop_shortest_distance}.
%The following result is a direct consequence of the remainders.

%\victor{en que momento usamos la sgte prop?}

\begin{proposition}\label{prop:bijection_vertical_dist}
%	Let $b=(b_1,b_2)\in \ZZ_2^+$ such that $\gcd(b_1,b_2)=1$. 
	For every $i\in \{1,\ldots,b_1-1\}$, the vertical distance from the point $(i,\lceil b_2i/b_1 \rceil)$ to the line segment $\calL$. 
%	satisfies $\lceil \frac{b_2}{b_1} i \rceil - \frac{b_2}{b_1} i $
	belongs to the set $\left\{1/b_1, 2/b_1, \ldots, (b_1 - 1)/b_1 \right\} $.
	Furthermore, for any $i\ne j$ in $\{1,\ldots,b_1-1\}$, the vertical distances from $(i,\lceil b_2i/b_1 \rceil)$ and $(j,\lceil b_2j/b_1 \rceil)$ to the line segment $\calL$ are different.
%	Furthermore, for each $i$, the vertical distance is a unique value in $\left\{\frac{1}{b_1}, \frac{2}{b_1}, \ldots, \frac{b_1 - 1}{b_1} \right\}$.
\end{proposition}

\begin{proof}
	Recall that we are assuming $\gcd(b_1,b_2)=1$.
	For every $i\in \{1,\ldots,b_1-1\}$ we have that
	$b_2 i = \left\lfloor b_2 i/b_1 \right\rfloor b_1 + \tau_i$
	where $\tau_i \in \{1,2,\ldots, b_1-1 \}$ is the remainder in the division.
	We remark that the remainders are non-zero since $b_1$ and $b_2$ are coprime. Therefore, the vertical distance from the point $(i,\lceil b_2i/b_1 \rceil)$ to $\calL$ is such that
	$\left\lceil b_2i/b_1  \right\rceil - b_2i/b_1  = (b_1-\tau_i)/b_1\in  \left\{1/b_1, 2/b_1, \ldots, (b_1 - 1)/b_1 \right\}$.
	One-to-one correspondence follows since the remainders are uniquely defined for the values $i\in \{1,2,\ldots,b_1-1\}$.
\end{proof}

%\victor{creo que no usamos la parte en azul }
\begin{proposition}
	\label{prop:vertical-distance}
Let $(x,b-x)$ be the partition computed by Algorithm~\ref{alg:two_colors_two_parts_algorithms}. 
Then, $x=(i^*, \lceil b_2 i^*/b_1 \rceil )$ is the point in $\calA_+$ that minimizes the vertical distance to the line segment $\calL$.
In particular, the vertical distance from $x$ to the line segment $\calL$ is $1/b_1$. 
%{\color{blue}Furthermore, the length of the $\ell_2$-projection from $x$ to the line segment $[0,b]$ is equal to $1/\|b\|_2$. }
%	The point $x$ returned by Algorithm~\ref{lem:characterization_gcd_eps} is the closest integer point in $\mathcal{A}_+$ to the line $0$-$b$. The vertical distance of $x$ to the line $0$-$b$ is $1/b_1$ while the $\ell_2$-distance to the line $0$-$b$ is $1/\|b\|_2$. 
\end{proposition}

\begin{proof}
	Recall that we are assuming $\gcd(b_1,b_2)=1$.
%	The result is direct if $\gcd(b)\geq 2$, since in that case $x\in [0,b]$. 
%	Suppose otherwise that $\gcd(b)=1$. 
	For each $i\in \{1,\ldots,b_1-1\}$, the closest point in $\mathcal{A}_+ \cap \{ (i, t) : t\in \ZZ_+ \}$ to the line segment $\calL$ is the point $(i, \lceil b_2i/b_1 \rceil)$ and the vertical distance from this point to $\calL$ corresponds to $\left\lceil b_2 i/b_1 \right\rceil - b_2i/b_1 = (b_1 - \tau_i)/b_1 \label{eq:vertical_dist}$,
	where $\tau_i\in \{1,\ldots, b_1-1\}$ is the remainder $\tau_i = b_2 \cdot i \pmod {b_1}= \tau \cdot i \pmod {b_1}$ and $\tau$ is the remainder $\tau= b_2 \pmod {b_1}$ . The minimum of $(b_1 - \tau_i)/b_1$ is attained when $\tau_i = b_1 - 1$, which is exactly what Algorithm~\ref{alg:two_colors_two_parts_algorithms} computes. 
	In particular, the vertical distance from this point $x$ to the segment is $1/b_1$. 
%	{\color{blue}The distance of the $\ell_2$-projection of the point $x$ 
	%$(i, \lceil \frac{b_2}{b_1} i \rceil)$ 
%	can be computed via similarity of triangles. 
%	See Figure~\ref{fig:prop_shortest_distance}.}
\end{proof}	
%	\begin{figure}[H]
%		\centering
%		\begin{tikzpicture}[scale=.6]
%			
%			
%			\foreach \x in {0,...,5}
%				\foreach \y in {0,...,7} 
%					\node [circle,fill=black,scale=0.23]  (\x\y) at (1*\x,1*\y) {}; 
%			
%			\node (-x) at (-0.5,0) {};
%			\node (x) at (5.5,0) {};
%			\node (y) at (0,7.5) {};
%			\node (-y) at (0,-0.5) {};
%			
%			\node(b1) at (5,0) [scale=0.1] {};
%			\node(b2) at (0,7) [scale=0.1] {};
%			\node(b) at (5,7) [scale=0.3, fill=black, circle] {};
%			\node(origin) at (0,0) [scale=0.1] {};
%			
%			
%			\node(z) at (3,5) [scale=0.3, fill=blue, circle] {};
%			
%			
%			
%			\path[draw=blue,every node/.style={sloped,anchor=south,auto=false}]
%			(origin) edge node {} (z)
%			(z) edge node {} (b)
%			;
%			
%			\path[draw=black,every node/.style={sloped,anchor=north,auto=false}]
%			(b) edge node {} (origin)
%			;
%			
%			%labels
%			\node at (b) [right] {$b$};
%			\node at (z) [left] {$x$};
%			\node (origen) at (-0.3,-0.3) {$0$};
%			
%			\path[dashed,thin,draw=gray]
%			(b1) edge node {} (b)
%			(b) edge node {} (b2)
%			;
%			
%			\path[red,thin]
%			(z) edge node {} (3,4.2)
%			(z) edge node {} (3.57,5)
%			(3.57,5) edge node {} (3,4.2)
%			;
%			
%			%axis
%			\path[draw,thick,<->] 
%			(-x) edge node {} (x)
%			(y) edge node {} (-y);
%			
%			%%
%			
%		\end{tikzpicture}
%		\caption{Depiction of points. The red triangle is similar to the triangle $0$-$b$-$b_1$.}
%		\label{fig:prop_shortest_distance}
%	\end{figure}

We are now ready to prove Theorem \ref{thm:two-two}.
%We strengthen the proposition by the following result which says that the point $x$ is not only the closest to the line, but also minimizes the slope among all possible solutions.

%\begin{proposition}
%	The point $x$ returned by Algorithm~\ref{alg:two_colors_two_parts_algorithms} minimizes $\min_{x\in \mathcal{A}_+} 1 - \frac{D(x)}{\gamma}$. Therefore, $(x,b-x)$ is an optimal solution of MRP with $2$ parts.
%\end{proposition}
	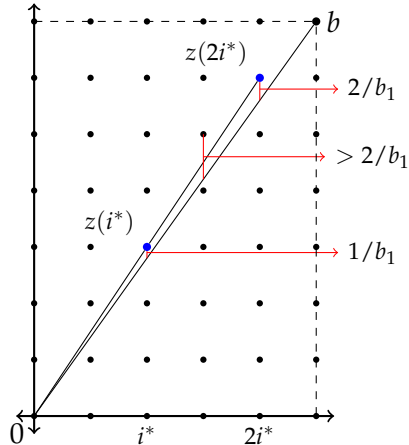
\begin{figure}[h!]
		\centering
		\begin{tikzpicture}[scale=.75]

			\foreach \x in {0,...,5}
			\foreach \y in {0,...,7} 
			\node [circle,fill=black,scale=0.23]  (\x\y) at (1*\x,1*\y) {}; 
			
			\node (-x) at (-0.5,0) {};
			\node (x) at (5.5,0) {};
			\node (y) at (0,7.5) {};
			\node (-y) at (0,-0.5) {};
			
			\node(b1) at (5,0) [scale=0.1] {};
			\node(b2) at (0,7) [scale=0.1] {};
			
			\node(b) at (5,7) [scale=0.3, fill=black, circle] {};
			\node(origin) at (0,0) [scale=0.1] {};
			\node(z) at (2,3) [scale=0.3, fill=blue, circle] {};
			\node(z1) at (4,6) [scale=0.3, fill=blue, circle] {};
			
			\node(p1) at (6,5.8) {\footnotesize$2/b_1$};
			\node(p0) at (6,2.9) {\footnotesize$1/b_1$};
			\node(p2) at (6,4.6) {\footnotesize$>2/b_1$};
			
			\path[draw=blue,every node/.style={sloped,anchor=south,auto=false}]
			(origin) edge node {} (z)
			(z) edge node {} (z1)
			;
			
			\path[draw=black,every node/.style={sloped,anchor=north,auto=false}]
			(b) edge node {} (origin)
			;
			
			\path[red,thin]
			(z1) edge node {} (4,5.6)
			(z) edge node {} (2,2.8)
			(3,5) edge node {} (3,4.2)
			;
			
			\path[red,thin,->]
			(4,5.8) edge node {} (p1)
			(2,2.9) edge node {} (p0)
			(3,4.6) edge node {} (p2)
			;
			
			%labels
			\node at (b) [right] {$b$};
			\node at (z) [above left=1pt] {\footnotesize$z(i^*)$};
			\node at (z1) [above left=1pt] {\footnotesize$z(2i^*)$};
			\node (origen) at (-0.3,-0.3) {$0$};
			\node at (20) [below] {\footnotesize$i^*$};
			\node at (40) [below] {\footnotesize$2i^*$};
			
			\path[dashed,thin,draw=gray]
			(b1) edge node {} (b)
			(b) edge node {} (b2)
			;

			%axis
			\path[draw,thick,<->] 
			(-x) edge node {} (x)
			(y) edge node {} (-y);

		\end{tikzpicture}
		\caption{Every point $\left(i,\left\lceil b_2i/b_1 \right\rceil\right)$ with $i\in (i^*,2i^*)$ has vertical distance to the segment $\calL$ of at least $3/b_1$.}
		\label{fig:prop_smallest_slope}
	\end{figure}

\begin{proof}[Proof of Theorem \ref{thm:two-two}]
	Observe that solving the $\PMD$ problem corresponds $\min_{y\in \calA_+}(1-D(y)/D(b))$, and by Proposition \ref{prop:technical-schur} \ref{schur-angle}, this can be written as  
	\[\min_{x\in \mathcal{A}_+} \Big(1 - \left(\cos \theta_x/\cos \theta_b \right)^2\Big).\] 
	This last problem is equivalent to $\max_{x\in \mathcal{A}_+} \cos \theta_x$, { which consists in finding $x\in \mathcal{A}_+$ of minimum slope. In other words, we look for $i\in \{1,2,\ldots,b_1-1\}$ that minimizes $ \left \lceil b_2i/b_1  \right\rceil/i$.} 
%	If $\gcd(b)\ge 2$ then we are done since $x\in [0,b]$.
%	Suppose that $\gcd(b)=1$.
	We show in what follows that the part $x$ constructed by Algorithm~\ref{alg:two_colors_two_parts_algorithms} solves this problem of minimum slope.
	
	Observe that $\left(\left \lceil b_2i/b_1 \right\rceil - b_2i/b_1\right)/i = \left \lceil b_2i/b_1 \right\rceil/i - b_2/b_1$ for every $i\in \{1,2,\ldots,b_1-1\}$. Therefore, the problem of minimizing the slope is equivalent to minimize the vertical distance from a point $\left(i, \left\lceil b_2i/b_1 \right\rceil \right)$ to the line segment $\calL$ normalized by $i$, with $i\in \{1,\ldots,b_1-1\}$.
	Let $i^*$ be the solution found by Algorithm~\ref{alg:two_colors_two_parts_algorithms} to the problem $i \cdot \tau= b_1 - 1 \pmod{b_1}$. 
	We study the area between the line that passes through the origin and $\left(i^*,\left\lceil b_2 i^*/b_1 \right\rceil\right)$ and the line $\calL$.
	% and we compare the vertical distance of this line to the line $0$-$b$. 
	Both of these lines can be explicitly parameterized by
	\[
	y(t) =  \frac{b_2}{b_1}t\quad \text{and} \quad
	z(t) = \left( \frac{1}{i^*} \left\lceil \frac{b_2}{b_1} i^* \right\rceil  \right)t,
	\]
	with $t\in [0,b_1]$.
	We define $\mathcal{L}_{i^*}= \{ (t,z(t)) : t \in [0,b_1] \}$ and observe that $\calL= \{ (t,y(t)) : t\in [0,b_1] \}$.
	Note that for every positive integer $\ell$ we have that
	\[
	z(\ell i^*) - y(\ell i^*) = \left( \frac{1}{i^*} \left\lceil \frac{b_2}{b_1} i^* \right\rceil - \frac{b_2}{b_1} \right) \ell \cdot i^* = \ell \left(\left\lceil \frac{b_2}{b_1} i^* \right\rceil - \frac{b_2}{b_1}i^* \right) = \frac{\ell}{b_1},
	\]
	where the last equality holds due to Proposition \ref{prop:vertical-distance}.
	That is, the vertical distance between two points in the two lines, sharing the same first coordinate, is increasing as multiples of $1/b_1$ when the first coordinate is an integer multiple of $i^*$. 
	We claim that in the interior of $\{ (t,w) : 0\leq t \leq b_1, y(t) \leq w \leq z(t) \}$ there are no integral points. 
%	Then, the interior of the area between the lines $\mathcal{L}_{i^*}$ and $\mathcal{L}_b$ between $i^*$ and $2i^*$ does not contain any integral point. This is because the vertical line of any point $(i, \lceil \frac{b_2}{b_1} i \rceil)$ to $\mathcal{L}_b$ is at least $2/b_1$ for $i\neq i^*$, $i=1,\ldots, b_1-1$. If $2i^* \leq b_1$, then the point $(2i^*,z(2i^*))$ corresponds to $\left(2i^*, \lceil \frac{b_2}{b_1} 2i^*\rceil \right)$ which is at vertical distance $2/b_1$ of $\mathcal{L}_b$. 
	Observe that for any positive integer $\ell$ such that $\ell \cdot i^* \leq b_1$ we have that the interior of the area between the line segments $\mathcal{L}_{i^*}$ and $\calL$ when the first coordinate lives in $(\ell i^*,(\ell+1)i^*)$ does not contain any integral point (see Figure~\ref{fig:prop_smallest_slope}). 
	{This is because the vertical distance between $(i, \lceil b_2i/b_1 \rceil)$ to $\calL$ is at least $(\ell+1)/b_1$ for $i\notin \{i^*,2i^*,\ldots,\ell i^*\}$ and $i\in \{1,\ldots, b_1-1\}$, which is true by Proposition~\ref{prop:bijection_vertical_dist}.} 
%	That concludes the claim and the proof of the theorem.
	{This concludes that among all possible points $\left( i, \left\lceil b_2i/b_1 \right\rceil  \right)$, the point $(i^*, z(i^*))$ achieves minimum slope which finishes the proof.}
\end{proof}

%
%\sebastian{asdasd}
%
%
%
%Recall that for $x\in\ZZ_+^2$ we define
%\[f(x) = \frac{\left(x+y\right)^2}{x^2+y^2} = 2\cos \theta_{(x,y)}\]
%Recall also that $f(b_1,b_2) = \gamma$. Therefore, the optimal partition satisfies
%\[\cos \theta_{(x,y)} = (1-\epsilon^*)\cos \theta_{(b_1,b_2)}\]
%This means that the best way to partition the balls corresponds to the vector with the closest angle to the original. Observe there is symmetry meaning line $(b_1,b_2)$ divides the area in two zones and for any partition one point lands in each zone. The optimal algorithm corresponds to compute (wlog assume $b_1\geq b_2$)
%\[\argmin_{x\in\{1,\ldots,\left\lfloor\frac{b_1}{2}\right\rfloor\}}\min\left\{\left\lceil\frac{b_2}{b_1}x\right\rceil-\frac{b_2}{b_1}x,\frac{b_2}{b_1}x - \left\lfloor\frac{b_2}{b_1}x\right\rfloor\right\}\]
%

\subsection{Analysis for $k$-Partitions}

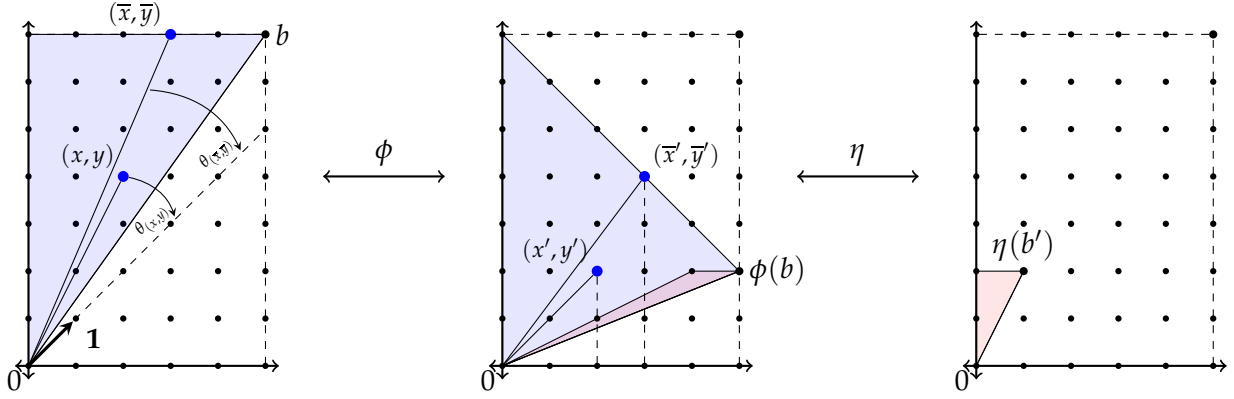
\begin{figure}[h!]
	\centering
	\begin{tikzpicture}[scale=0.63]

		\foreach \x in {0,...,5}
		\foreach \y in {0,...,7} 
		\node [circle,fill=black,scale=0.23]  (\x\y) at (1*\x,1*\y) {}; 
		
		\draw [fill=blue, fill opacity=0.1] (0,0) -- (5,7) -- (0,7) -- cycle;
		
		\node (-x) at (-0.5,0) {};
		\node (x) at (5.5,0) {};
		\node (y) at (0,7.5) {};
		\node (-y) at (0,-0.5) {};
		
		\node(b1) at (5,0) [scale=0.1] {};
		\node(b2) at (0,7) [scale=0.1] {};
		\node(b) at (5,7) [scale=0.3, fill=black, circle] {};
		\node(origin) at (0,0) [scale=0.1] {};
		
		\node(ONE) at (1,1) [scale=0.1] {};
		
		\node(z) at (2,4) [scale=0.4, fill=blue, circle] {};
		\node(zz) at (3,7) [scale=0.4, fill=blue, circle] {};

		\path[draw=blue,every node/.style={sloped,anchor=south,auto=false}]
		(origin) edge node {} (z)
		(origin) edge node {} (zz)
		;
		
		\path[draw=black,every node/.style={sloped,anchor=north,auto=false}]
		(b) edge node {} (origin)
		;
		
		%labels
		\node at (b) [right] {$b$};
		\node at (z) [above left,scale=0.8] {$(x,y)$};
		\node at (zz) [above left,scale=0.8] {$(\overline{x},\overline{y})$};
		\node (origen) at (-0.3,-0.3) {$0$};
		\node at (ONE) [below right] {$\mathbf{1}$};
		
		\path[dashed,thin,draw=gray]
		(b1) edge node {} (b)
		(b) edge node {} (b2)
		(origin) edge node {} (5,5)
		;
		
		%		\path[red,thin]
		%		(z) edge node {} (3,4.2)
		%		(z) edge node {} (3.57,5)
		%		(3.57,5) edge node {} (3,4.2)
		%		;
		
		%axis
		\path[draw,thick,<->] 
		(-x) edge node {} (x)
		(y) edge node {} (-y);
		
		\path[draw,very thick,->,>=stealth,shorten >=1pt]
		(origin) edge node {} (ONE);
		
		\path[draw, ->,>=stealth, shorten >=1pt]
		(z) edge [bend left=30] node [below left, rotate=45,scale=0.6] {$\scriptsize\theta_{(x,y)}$} (3.1,3.1)
		;
		\node(inzz) at (2.5,5.83) [scale=0.3] {};
		\draw[->,>=stealth, shorten >=1pt]
		(inzz) edge [bend left=30] node [below=13pt, rotate=45,scale=0.6] {$\scriptsize\theta_{(\overline{x},\overline{y})}$} (4.5,4.5)
		;

		\node(LEFT) at (6,4) {};
		
		\begin{scope}[xshift=10cm]
			
			\node(RIGHT) at (-1,4) {};
			
			\draw [fill=blue, fill opacity=0.1] (0,0) -- (5,2) -- (0,7) -- cycle;
			\draw [fill=red, fill opacity=0.1] (0,0) -- (5,2) -- (4,2) -- cycle;
			
			\foreach \x in {0,...,5}
			\foreach \y in {0,...,7} 
			\node [circle,fill=black,scale=0.23]  (\x\y) at (1*\x,1*\y) {}; 
			
			\node (-x) at (-0.5,0) {};
			\node (x) at (5.5,0) {};
			\node (y) at (0,7.5) {};
			\node (-y) at (0,-0.5) {};
			
			\node(b1) at (5,0) [scale=0.1] {};
			\node(b2) at (0,7) [scale=0.1] {};
			\node(b) at (5,7) [scale=0.3, fill=black, circle] {};
			\node(bp) at (5,2) [scale=0.3, fill=black, circle] {};
			\node(origin) at (0,0) [scale=0.1] {};

			\node(z) at (2,2) [scale=0.4, fill=blue, circle] {};
			\node(zz) at (3,4) [scale=0.4, fill=blue, circle] {};

			\path[draw=blue,every node/.style={sloped,anchor=south,auto=false}]
			(origin) edge node {} (z)
			(origin) edge node {} (zz)
			;
			
			\path[draw=black,every node/.style={sloped,anchor=north,auto=false}]
			(bp) edge node {} (origin)
			;
			
			%labels
			\node at (bp) [right] {$\phi(b)$};
			\node at (z) [above left,scale=0.8] {$(x',y')$};
			\node at (zz) [above right,scale=0.8] {$(\overline{x}',\overline{y}')$};
			\node (origen) at (-0.3,-0.3) {$0$};
			
			\path[dashed,thin,draw=gray]
			(b1) edge node {} (b)
			(b) edge node {} (b2)
			(z) edge node {} (2,0)
			(zz) edge node {} (3,0)
			;
			
			%		\path[red,thin]
			%		(z) edge node {} (3,4.2)
			%		(z) edge node {} (3.57,5)
			%		(3.57,5) edge node {} (3,4.2)
			%		;
			
			%axis
			\path[draw,thick,<->] 
			(-x) edge node {} (x)
			(y) edge node {} (-y);
			
			\node(LEFT2) at (6,4) {};
		\end{scope}

		\path[draw,thick, <->]
		(LEFT) edge node[above] {$\phi$} (RIGHT);
	
		\begin{scope}[xshift=20cm]
			
			\node(RIGHT2) at (-1,4) {};
			
			\draw [fill=red, fill opacity=0.1] (0,0) -- (1,2) -- (0,2) -- cycle;
			
			\foreach \x in {0,...,5}
			\foreach \y in {0,...,7} 
			\node [circle,fill=black,scale=0.23]  (\x\y) at (1*\x,1*\y) {}; 
			
			\node (-x) at (-0.5,0) {};
			\node (x) at (5.5,0) {};
			\node (y) at (0,7.5) {};
			\node (-y) at (0,-0.5) {};
			
			\node(b1) at (5,0) [scale=0.1] {};
			\node(b2) at (0,7) [scale=0.1] {};
			\node(b) at (5,7) [scale=0.3, fill=black, circle] {};
			\node(bp) at (1,2) [scale=0.3, fill=black, circle] {};
			\node(origin) at (0,0) [scale=0.1] {};

			\path[draw=black,every node/.style={sloped,anchor=north,auto=false}]
			(bp) edge node {} (origin)
			;
			
			%labels
			\node at (bp) [above] {$\eta(b')$};
			\node (origen) at (-0.3,-0.3) {$0$};
			
			\path[dashed,thin,draw=gray]
			(b1) edge node {} (b)
			(b) edge node {} (b2)
			;
			
			%		\path[red,thin]
			%		(z) edge node {} (3,4.2)
			%		(z) edge node {} (3.57,5)
			%		(3.57,5) edge node {} (3,4.2)
			%		;
			
			%axis
			\path[draw,thick,<->] 
			(-x) edge node {} (x)
			(y) edge node {} (-y);

		\end{scope}
			\path[draw,thick, <->]
			(LEFT2) edge node[above] {$\eta$} (RIGHT2);
		
	\end{tikzpicture}
	\caption{Transformations $\phi_{b}$ and $\eta_{b'}$, where the subindices are ignored. In this figure, we have $b= (5,7)$, $\kappa =1$, $b' = (5,2)$ and $m = 2$. Note that if the angle $\theta_{(x,y)}$ is smaller than the angle $\theta_{(\overline{x},\overline{y})}$, then the slopes of $(x',y')=\phi(x,y)$ is smaller than the slope of $(\overline{x}',\overline{y}')=\phi(\overline{x},\overline{y})$, and the converse also holds. A similar order invariant property is held by $\eta$ in the red area.}
	\label{fig:transformation_phi}
\end{figure}

In this section, we present an algorithm for the general $\PMD$ problem with any $k\geq 3$ and two types. Given a budget $(b_1,b_2)$, Algorithm~\ref{alg:two_colors_k_parts_good} recursively reduces the instance size until reaching an initial condition where a pattern-like solution of $k$ parts can be easily computed (Lemma~\ref{lem:optimal_k_parts}). 
In what follows, we assume that $b_1$ does not divide $b_2$ since otherwise we can handle it by Theorem~\ref{lem:characterization_gcd_eps}. We denote by $\mathrm{slope}(x,y)= y/x$ the slope of the point $(x,y)$. Let 
%$\kappa= \max\{ i \in \ZZ : i \cdot b_1 \leq b_2 \}$. 
$\kappa= \lfloor b_2/b_1 \rfloor$ and
define $b_1'=b_1$ and $b_2'=b_2-\kappa b_1$. 
Let 
%$m= \max\{ i\in \ZZ : i \cdot b_2' \leq b_1'  \}$. 
$m= \lfloor b_1'/b_2'\rfloor$. 
We consider two functions $\phi_{b}$ and $\eta_{b'}$ defined as follows. Function $\phi_{b}$ maps a point $(x,y)\in \mathcal{A}_+=  \{ (u,v)\in \mathcal{A}: v \geq (b_2/b_1)u   \}$ to the point $(x,y- \kappa x)$. 
Consider 
\begin{equation*}
\mathcal{B}_+^{b}= \phi_b(\calA_+)\cap \Big\{ (u,v): v\leq b_2'  \Big\}\cap \Big\{  (u,v) :  v\leq ( 1- m b_2'/b_1' )u  \Big\}. 
\end{equation*}
Function $\eta_{b'}$ maps $(x,y)\in \calB_+^b$ to the point $(x-m y, y)$. A depiction of these two functions appear in Figure~\ref{fig:transformation_phi}.
The following proposition summarizes the main properties of the functions $\phi_b$ and $\eta_{b'}$.

\begin{proposition}\label{prop:functions_properties}
	Consider a budget vector $(b_1,b_2)$.
	Then, the following holds:
	\begin{enumerate}[label=(\alph*)]
		\item Functions $\phi_b: \mathcal{A}_+ \to \phi_b(\calA_+)$ and $\eta_{b'} : \calB_+^b \to  \eta_{b'}(\calB_+^b)$ are one-to-one. \label{function-a}
		
		\item For any two points $(x,y),(\overline{x},\overline{y})\in \mathcal{A}_+$, the slope between $\phi(x,y)$ and $\phi(\overline{x},\overline{y})$ is equal to \[\frac{y-\overline{y}}{x-\overline{x}}-\kappa.\] Moreover, $D(x,y)\geq D(\overline{x}, \overline{y})$ if and only if $\mathrm{slope}\left(\phi_b(x,y)\right)\leq \mathrm{slope}\left( \phi_b(\overline{x},\overline{y}) \right)$.\label{function-b}		
		
		\item For any two points $(x,y),(\overline{x},\overline{y})\in \calB_+^b$, the slope between $\eta_{b'}(x,y)$ and $\eta_{b'}(\overline{x},\overline{y})$ is equal to \[\frac{y-\overline{y}}{x-\overline{x}- m(y-\overline{y})}.\] Moreover, $\mathrm{slope}(x,y)\leq \mathrm{slope}(\overline{x},\overline{y})$ if and only if $\mathrm{slope}\left(\eta_{b'}(x,y)\right) \leq \mathrm{slope} \left(\eta_{b'}(\overline{x},\overline{y})\right)$. \label{function-c}
	\end{enumerate}
\end{proposition}

\begin{proof}
We have that \ref{function-a} holds directly since $\phi_b$ and $\eta_{b'}$ are linear transformation defined by non-singular matrices.
The first part of \ref{function-b} holds by directly computing the slope between the two points $(x,y-\kappa x)$ and $(\overline{x},\overline{y}-\kappa \overline{x})$. 

Note that the slope of the vector $\phi_b(x,y)\in\phi_b (\mathcal{A})$ equals $y/x- \kappa$, which is a translation of the slope of $(x,y)$.
Therefore, $\mathrm{slope}(x,y)\leq \mathrm{slope}(\overline{x},\overline{y})$ if and only if $\mathrm{slope}(\phi_{b}(x,y))\leq \mathrm{slope}(\phi_b(\overline{x},\overline{y}))$. The angle $\theta_{(x,y)}$ formed between $(x,y)$ and $\mathbf{1}$ also holds $\theta_{(x,y)}\leq \theta_{(\overline{x},\overline{y})}$ if and only if $\mathrm{slope}(x,y)\leq \mathrm{slope}(\overline{x},\overline{y})$. Since $D(x,y)=2 (\cos \theta_{(x,y)})^2$, and cosine is decreasing for $\theta_{(x,y)}\in [0,\pi/2]$, we conclude \ref{function-b}.
The proof of part \ref{function-c} is analogous to the previous point and the conclusion follows by the monotonicity of the function $f(s)= s/(1-sm)$, where $s=(y-\overline{y})/(x-\overline{x})$. 
\end{proof}

%For points $(x,y),(\overline{x},\overline{y})\in \calA_+$, we refer to the property $D(x,y)\geq D(\overline{x}, \overline{y})$ if and only if $\mathrm{slope}\left(\phi_b(x,y)\right)\leq \mathrm{slope}\left( \phi_b(\overline{x},\overline{y}) \right)$ as \emph{$\phi$-order invariance}, and we say that $\phi_b$ is \emph{$\phi$-order invariant}. Likewise, for points $(x,y)(\overline{x})\in \calB_+^b$, we refer to the property $\mathrm{slope}(x,y)\leq \mathrm{slope}(\overline{x},\overline{y})$ if and only if $\mathrm{slope}\left(\eta_{b'}(x,y)\right) \leq \mathrm{slope} \left(\eta_{b'}(\overline{x},\overline{y})\right)$ as \emph{$\eta$-order invariance}, and we say that $\eta_{b'}$ is \emph{$\eta$-order invariant}.

\begin{algorithm}[h!]
	%	\SetAlgoNoLine
	\caption{Two types and $k$ parts}\label{alg:two_colors_k_parts_good}
	\begin{algorithmic}[1]
		\Require{Budget $b = (b_1,b_2)$ with $b_1\leq b_2$ and $k\geq 3$ parts. }
		\Ensure{A set of $k-1$ breakpoints $w_{ 1},\ldots,w_{ k-1}$.}
		\State Compute $\kappa$, $b'=(b_1',b_2')$, $m$ and functions $\phi_{b}$ and $\eta_{b'}$. 
		\If {$b_1' - \ell b_2' + 1 \leq k \leq b_1' - (\ell - 1)b_2'$ for some $\ell\in \{1,\ldots,m\}$}
		\State Construct points $(w_1',\ldots,w_{k-1}')$ as follows: 
		\State \multiline{Select the $b_2'$ points given by $w_{i-1}'=(i \ell, i)$ for $i\in \{1,\ldots,b_2'\}$ and complete the rest with any $k-b_2'-1$ points on the line joining $(\ell b_2',b_2')$ and $b'$.} 
		\Else { $k \leq b_1' - m b_2'$}
		\State Recur on the input $\left(\eta_{b'}(b'),k\right)$ and obtain points $(w_1'',\ldots,w_{k-1}'')$. 
		\State Define $(w_1',\ldots,w_{k-1}')=(\eta_{b'}^{-1}(w_1''), \ldots, \eta_{b'}^{-1}(w_{k-1}''))$.
		\EndIf
		\State Return $(w_1,\ldots,w_{k-1})=(\phi_b^{-1}(w_{1}'), \ldots, \phi_b^{-1}(w_{k-1}'))$.
	\end{algorithmic}
\end{algorithm}

We give a brief explanation of the role of the functions $\phi_b$ and $\eta_{b'}$. Any optimal $k$ segments joined by $k-1$ breakpoints in $\mathcal{A}_+$ can be reordered by decreasing slope. Thus the diversity of the first segment corresponds to the diversity of the $k$ parts (Lemma~\ref{lem:there_is_sol_decreasing_slope}). 
By mapping the $k-1$ points using transformation $\phi_b$ and using Proposition~\ref{prop:functions_properties}, we see that the $k$ segments joining these new points are ordered by decreasing slope. The converse is also true by the same argument. If the number of segments required is larger than $b_1'-mb_2'$, then the slope of any $k$ segments joining $0$ with $b'$ must have a slope of at least $1/\ell$ for an appropriate $\ell$. We are able to explicitly construct a solution with slope $1/\ell$ which can be mapped back to a solution of the $k$-partition problem using $\phi_b^{-1}$. If the number of parts required $k$ is at most $b_1'-mb_2'$, then, intuitively, pairs of points with large slope can be discarded, namely points in $\phi_b(\calA_+) \setminus \calB_+^b$. Since the slopes are preserved under $\eta_{b'}$ due to $\eta$-order invariance, we can recur over the instance $\eta_{b'}(\calB_+^b)$. 

The points found in the recursion can be brought back to the initial instance using $\eta_{b'}^{-1}$ and $\phi_{b}$ and using their corresponding order invariance. This is formally presented in Algorithm~\ref{alg:two_colors_k_parts_good}. We describe the algorithm for $k\geq 3$ since $k=2$ is already solved by Algorithm~\ref{alg:two_colors_two_parts_algorithms}.
%
%The next proposition shows that the slope formed by any two points in the space $\mathcal{A}_+$ is preserved, up to a constant, after mapping these two points via $\phi$. It also shows that diversity in the space $\mathcal{A}_+$ is in correspondence with the slope in the space $\mathcal{A}_+'$.
%
% joining $\mathcal{L}$ (see Algorithm~\ref{alg:two_colors_k_parts}). The correctness of this method is proven in Theorem~\ref{thm:two-k}. Intuitively, the slopes formed by these $k$ segments are the best approximations for the slope of the line $\mathcal{L}$. This combined with the fact that slope and diversity are closely related, then we obtain the correctness of our algorithm. This is formally proven in Theorem~\ref{thm:optimal_k_parts}. A naive implementation of this method can be done in time $\mathcal{O}(k \max\{ b_1,b_2\})$ by searching iteratively the $k-1$ closest points to $\mathcal{L}$. We present an implementation that takes time $\mathcal{O}(k (\log \max\{ b_1,b_2\})^2)$. 
%\alfredo{maybe put algorithm with if and else?}
From the breakpoints $w_1,\ldots,w_{k-1}$ computed by the algorithm, we recover a $k$-partition by defining $x_1=w_1$, $x_k=b-w_{k-1}$ and 
\begin{equation}
	x_j=w_j-w_{j-1} \label{eq:k-partition}
\end{equation}
for every $j\in \{2,\ldots,k-1\}$. Note that the algorithm is well-defined. Indeed, by the choice of $\kappa$ and $m$, in each recursive call we always have $b_1\leq b_2$. In each iteration $b_1$ and $b_2$ decrease, so given that $k\geq 3$, there must be a recursive call where the corresponding $b_1'$ and $b_2'$ hold $b_1' - \ell b_2' + 1 \leq k \leq b_1' - (\ell - 1)b_2'$ for some $\ell\in \{1,\ldots,m\}$.

Note that the maximum number of calls is bounded by the number of times that takes to reach $(b_1,b_2)=(1,1)$. Due to the implementation of the algorithm, we can see that this is at most $\mathcal{O}(\log \max\{ b_1, b_2 \})$ recursive calls. Thus, the overall number of operations is $\mathcal{O}(k \log \max\{ b_1, b_2\})$. Including the time of arithmetic operations, we see that the time complexity is increased by at most $\mathcal{O}(\log \max\{b_1,b_2\})$ factor, which gives us an algorithm with overall time complexity 
\[\mathcal{O}(k\log^2 \max\{ b_1, b_2\}),\] 
that is polynomial in the input $(b_1,b_2),k$, and the output length, a vector of length $k$.
The following theorem summarizes our main result in this section.

\begin{theorem}
	\label{thm:two-k}	
	For every budget $b=(b_1,b_2)$ with $\gcd(b)=1$ and $b_1\le b_2$, the $k$-partition $x_1,\ldots,x_k$ in \eqref{eq:k-partition} obtained from Algorithm~\ref{alg:two_colors_k_parts_good} solves the $\PMD$ problem with $k\geq 3$ parts. 
	The algorithm runs in time $\mathcal{O}(k\log^2 \max\{ b_1, b_2\})$.
	%	In particular, in this case we have that $\varepsilon(b,2) = 1 - \frac{D(x)}{D(b)}$.
\end{theorem}

The proof of the theorem is a consequence of the following two structural results.

\begin{lemma}\label{lem:there_is_sol_decreasing_slope}
	For $b_1\leq b_2$ and $k\geq 2$, there is a $k$-partition solution $x_1,\ldots,x_k$ of $\PMD$ described by points $w_0=0, w_1,\ldots,w_{k-1}, w_k=b$ as $x_i = w_i - w_{i-1}$ for $i\in [k]$ and such that the slopes of the segments $w_{i-1}$-$w_{i}$ are decreasing and $\min_{i\in [k]} D(x_i) = D(w_1)$.
\end{lemma}

%and

\begin{lemma}\label{lem:optimal_k_parts}
	Let $(b_1,b_2)$ with $b_1\leq b_2$ and let $k\geq 3$. Suppose that for some $\ell \in [m]$ we have $b_1' - \ell b_2' + 1 \leq k \leq b_1' - (\ell - 1)b_2'$, where $b_1'=b_1$ and $b_2'=b_2-\lfloor b_2/b_1\rfloor b_1$. Then, any $k-1$ points in $\phi_b(\calA_+)$ will have a segment joining two points with a slope at least $1/\ell$.
\end{lemma}

Before we provide the proofs of the lemmata, we conclude Theorem~\ref{thm:two-k}.

\begin{proof}[Proof of Theorem~\ref{thm:two-k}.]
	 Since optimality is preserved under $\phi_b$ and $\eta_{b'}$, it is enough to show the result for one level of the recursion. Assume that for some $\ell \in [m]$ we have $b_1' - \ell b_2' + 1 \leq k \leq b_1' - (\ell - 1)b_2'$. Take the optimal solution $w_1,\ldots,w_{k-1}\in \calA_+$ given by Lemma~\ref{lem:there_is_sol_decreasing_slope}. Map these points using $\phi_b$: $\widetilde{w}_i=\phi_b(w_i)$ for $i \in [k-1]$ and let $\widetilde{w}_0=0$ and $\widetilde{w}_k=\phi_b(b)$. By Lemma~\ref{lem:optimal_k_parts}, we know that the largest slope of the $k$ segments formed by the points $\widetilde{w}_0,\ldots,\widetilde{w}_k'$ is at least $1/\ell$ and the slopes are sorted in decreasing order by Proposition~\ref{prop:functions_properties}. Now consider the solution constructed by Algorithm~\ref{alg:two_colors_k_parts_good}, namely $(w_1',\ldots,w_{k-1}')$. Since the largest slope of this solution is exactly $1/\ell$, we have $\mathrm{slope}(w_1') = 1/\ell \leq \mathrm{slope}(\widetilde{w}_1)$. Using Proposition~\ref{prop:functions_properties} again we obtain $D(\phi_b^{-1}(w_1)) \geq D(w_1)$, which concludes the optimality of the solution provided by Algorithm~\ref{alg:two_colors_k_parts_good}.
\end{proof}

\begin{proof}[Proof of Lemma~\ref{lem:there_is_sol_decreasing_slope}]
	Consider an optimal solution $x_1,\ldots,x_k\in \ZZ_+^2$. Since these points are vectors in $\RR^2$ we can compute their slopes $\mathrm{slope}(x_i)$. Without loss of generality, we assume that the points $x_1,\ldots,x_k$ are sorted by decreasing slopes: $\mathrm{slope}(x_1) \geq \cdots \geq \mathrm{slope}(x_k)$. Define $w_0=0$ and $w_i= w_{i-1}+ x_i$ for $i\in [k]$. Thus $w_k=b$. 
	
	We first claim that $\min_{i\in [k]} D(x_i) = \min \{ D(x_1), D(x_k)  \}$. Indeed, let $\alpha_i$ be the angle formed by $x_i$ and $(1,0)$. Then, by the order of $x_i$ we have $\alpha_1 \geq \ldots \geq \alpha_k$. Define $\widetilde{\theta}_i$ for $i\in [k]$ as follows: $\widetilde{\theta}_i=\theta_{x_i}$ if $x_i$ is above the line $\{(t,t):t\in \RR\}$ and $\widetilde{\theta}_i= -\theta_{x_i}$ otherwise. Thus $\theta_i + \pi/4= \alpha_i$ for all $i$. By the monotonicity of $\alpha_i$ we have $\pi/4\geq \theta_1 \geq \cdots \geq \theta_i\geq -\pi/4$. Since $\cos(x)$ is concave for $x\in [-\pi/2,\pi/2]$ and even\footnote{A function $f$ over $\RR$ is even if $f(x)= f(-x)$ for every $x\in \RR$.}, we have that the minimum of $\cos(x)$ for $x\in \{  \theta_{x_1} ,\ldots, \theta_{x_k} \}$ must be attained at $x\in \{\theta_{x_1}, \theta_{x_k}\}$. Using $D(x_i) = 2 (\cos \theta_{x_i})^2$, the result follows.

	We now show that the diversity is defined just by $x_1$, and if not, then we can modify slightly the solution $x_1,\ldots,x_k$ to achieve this. Note that if $\theta_k \geq 0$, with $\theta_k$ defined as before, then the result follows by the monotonicity of the cosine function. Suppose then that $\theta_{k} < 0$. Let $\calL'=\{ (t, \kappa \cdot t +\tau) : 0\leq  t \leq b_1  \}$ be the continuous line joining $(0,\kappa)$ and $b=(b_1,b_2)$, where $\tau= b_2 \pmod{b_1}$ and $\kappa = \lfloor b_2/b_1 \rfloor$. Thus the point $w_{k-1}$ lies above the line $\calL'$. Let $i^*$ be the first index where $w_{i^*}$ is below $\calL'$ and $w_{i^*+1}$ is above the line $\calL'$. Without loss of generality $w_{i^*} = (p,q)$, thus, the first component of $w_{i^*+1}$ is at least $p+1$. This implies that $k-1-i^*$, the number of points $w_{i^*+1},\ldots, w_{k-1}$, is at most $b_1-1-p$. Consider the following solution: $w_i' = w_i$ for $i\leq i^*$ and $w_{i^*+1}' = ( p+1, \kappa(p+1) + \tau   ) , \ldots, w_{k-1}' = ( p + \ell, \kappa(p+\ell) + \tau ) $, where $\ell= k-1-i^*$, and $w_k'=b$. Then, we observe that the parts $x_i'= w_i'-w_{i-1}'$ for $i\in [k]$ exhibit a diversity as good as the diversity of $x_1,\ldots,x_k$, the slopes of $x_i'$ are in decreasing order, and all $x_i'$ lie above or in the line $\{(t,t):t\in \RR\}$, which ensures that their corresponding angle $\theta_i'\geq 0$ for all $i$.  
\end{proof}

\begin{proof}[Proof of Lemma~\ref{lem:optimal_k_parts}]
	By contradiction, assume there is a solution of $k-1$ points where the $k$ segments have slope smaller than $1/\ell$. Take any of these segments and suppose that $(x,y)$ and $(\overline{x}, \overline{y})$ are its endpoints, then the ratio between $\Delta y = y-\overline{y}$ and $\Delta x= x-\overline{x}$ is strictly smaller than $1/\ell$. Let us assume without loss of generality that $x\geq\overline{x}$. Since the points $x,\overline{x},y$ and $\overline{y}$ are integer and $\ell$ is also an integer, we deduce
	\[
	\Delta x\geq 1 + \ell \cdot \Delta y.
	\]
	Adding up the $\Delta x$'s of all segments, we obtain
	\begin{align*}
		b_1' & = \sum_{\text{segments}} \Delta x \geq \sum_{\text{segments}} (1+\ell \cdot \Delta y)= k+\ell b_2',
	\end{align*}
	but this contradicts the assumption $b_1' \leq \ell b_2' +k - 1$. Thus, in any set of $k-1$ points in $\phi_b(\calA_+)$ there must be a pair of points where the segment joining them has slope of at least $1/\ell$.
\end{proof}

\section{Challenges in Higher Dimension and Open Questions}\label{sec:challenges}

In this section, we provide a brief insight on the challenges that our problem poses when dealing with more than two types and we discuss some of the remaining open questions. First, for $r=2$, the two dimensional geometry makes the problem more approachable in that we are able to narrow the search and focus on the nearest points around the line segment $\calL$ determined by $(b_1,b_2)$. This approach works for both the case $k=2$ and the recursive procedure for $k\geq 3$. This is possible since the diversity $D(x)$ of a vector $x$, the angle $\theta_x$ between $x$ and $\mathbf{1}$, and the slope of $x$ are directly connected. Moreover, during the analysis of $k=2$ parts, we uncovered that the distances of the closest points are in one-to-one correspondence with the set $\{1,\ldots,\tau\}$, where $\tau=b_2 \pmod{b_1}$. Hence the closest point to the line $\calL$ defines a $2$-partition of maximin diversity.

Now, we discuss the geometric challenges in higher dimension ($r\geq3$). For simplicity, consider the $2$-partition problem. 
The approaches mentioned above are not directly applicable to this case. 
First, note that for dimensions larger than $2$, we lose the notion of the slope of a vector $x$, which is crucial for the recursion in Algorithm~\ref{alg:two_colors_k_parts_good}. %and for solving the case of at least $3$ parts. 
Furthermore, for $r\geq 3$, it is possible to construct a counterexample that shows that the closest point (with respect to the Euclidean distance) to the line segment $\calL=\{t\cdot b: \ t\in[0,1]\}$---approach used for $r=2$ in Algorithm~\ref{alg:two_colors_two_parts_algorithms}---does not define an optimal solution of the $\PMD$ problem for $k=2$.

In higher dimensions, we need a different geometric perspective of the problem.
Observe that the rotation of $b$ around $\mathbf{1}$ defines a cone, whose boundary is composed by points in $\RR_+^r$ with the same diversity index as $b$.
Let \[\calK_b = \Big\{y\in \RR_+^r: \ \|y\|_1^2\geq D(b)\|y\|_2^2\Big\}\] be the quadratic cone determined by the global diversity $D(b)$. 
Note that $D(x)\geq D(b)$ if and only if $x\in\calK_b$. 
Therefore, a tentative approach is to look for integral points outside $\calK_b$ of minimum Euclidean distance, which corresponds to a projection problem. However, we need to solve the following question: are the integral points (outside $\calK_b$) of minimum projection in one-to-one correspondence with the points of maximum diversity? How many of these points exist and which one do we choose?  The first question can be posed as addressing the existence of integral points between two quadratic cones, which also may be related to open questions in number theory.

A second approach is similar to the case $r=2$: to look for a piecewise linear approximation of the segment $\calL$ with a breakpoint in $x$. 
Intuitively, the point $x$ should have a small angle with respect to $\calL$ and with respect to $\mathbf{1}$. 
Indeed, if this is not the case, then the angles with respect to $\mathbf{1}$ defined by each part , namely $\theta_x$ and $\theta_{b-x}$, may increase which gives a worse diversity objective; recall that the objective for the $2$-partition problem can be equivalently written as $2\cdot\min\{\cos^2(\theta_x),\cos^2(\theta_{b-x})\}$. 
To find this breakpoint $x$, a tentative idea is to enumerate the points that belong to the integral hypercube covering of $\calL$. However, it is not immediate to show that the points outside this cover have a worse objective.

%Note that any $2$-partition $(x,b-x)$ with $x\in\ZZ_+^r$ must satisfy that the line between $x$ and $b-x$ passes through $b/2$. 
One of the main remaining open questions relates to the complexity of the general problem. It is not clear whether or not a polynomial time algorithm exists for $2$-partitions with $r>2$.
When the number of types $r$ is constant, we conjecture that for $2$-partitions a pseudo-polynomial time algorithm exists. Tentative approaches are: (1) an appropriate enumeration of feasible points or (2) by using the problem of minimum projection over the cone $\calK_b$. However, the general $\PMD$ problem might no be solvable in polynomial time. 
This question aligns with the open question stated by Laber et al. \cite{laber2018binary}, where the complexity of the problem remains unanswered for the Gini impurity measure.
Also, the approach taken in \cite{laber2018binary} might be useful to design algorithms with constant approximation guarantees for our problem.

Another open question corresponds to the scarce resource setting where $k>b_j$ for some $j\in [r]$. In this case, the diversity of the resulting partition might be considerably worse than the global diversity, since some subgroups do not get individuals of type $j$. Finally, it would be interesting to study our framework under other diversity indices, such as the general class of Hill numbers or similarity-based indices \cite{hill1973diversity,daly2018ecological}. Formally, for $x\in\ZZ^r_+$, the class of Hill numbers is defined as
\[
D_q(x) = \left(\sum_{i\in[r]}\frac{x_i^q}{\|x\|^q_1}\right)^{\frac{1}{1-q}},
\]
where $q\in[0,\infty]$ is known as the \emph{order} of the diversity. For $q= 2$, we recover the Simpson dominance index. For $q=1$, the limit exists and corresponds to the exponential of the Shannon entropy. For $q=\infty$, the metric measures only the maximum of the entries. The Hill numbers are the only family of ecological diversity metrics that are known to satisfy key mathematical axioms \cite{daly2018ecological}. For $q\neq 2$, our proof techniques are not directly applicable since in this case the metric may not have a geometric interpretation, in particular Property (b) in Proposition \ref{prop:technical-schur} does not apply.

% \sebastian{although our proof technique are not directly applicable as other indices cannot be interpreted as angles between vectors.}

%\input{multiple_types}

\section{Conclusions}\label{sec:conclusions}

This work presented a novel framework for the partition problem under diversity requirements. We provide a geometric interpretation of the relationship between the global diversity of the community and each subgroup's diversity. We show that a perfect partition exists only when the number of parts $k$ divides the gcd of $b$. We also design a polynomial time algorithm for the case of $r=2$ types. Finally, we discuss the technical challenges that we face in higher dimensions and some open questions.

Addressing diversity concerns has posed numerous challenges. Long-term interdisciplinary efforts and multiple views on the matter are needed to appropriately progress towards fairness and equity \cite{chi2021reconfiguring,bernstein2020diversity}. We hope that, from a technical perspective, our work helps in understanding the effects on diversity when dividing a community into subgroups. We think that our framework and results could provide deeper insights in other resource constrained settings that look to incorporate diversity requirements such as clustering, classification and scheduling.

\bibliographystyle{abbrv}
{\small \bibliography{bibliography_diverse}}
%\appendix

%\input{appendix}
\end{document}